\documentclass[11pt]{EMPIR_GPG}

\guideversion{1.0}
\guidedate{\today}

\usepackage[english]{babel}
\usepackage{booktabs}
\usepackage{hyphenat}
\usepackage{longtable}
\usepackage{graphicx}
\usepackage{siunitx}
\usepackage{csquotes}
\usepackage{sfmath}
\usepackage{subcaption}
\usepackage{pdfpages}

\DeclareSIUnit \bar {bar}
\graphicspath{{./images/}}

\title{Good practice guide on the graphene-based \\ AC-QHE realization of the farad}
\date{}
\author{}

\begin{document}

\includepdf[pages=1]{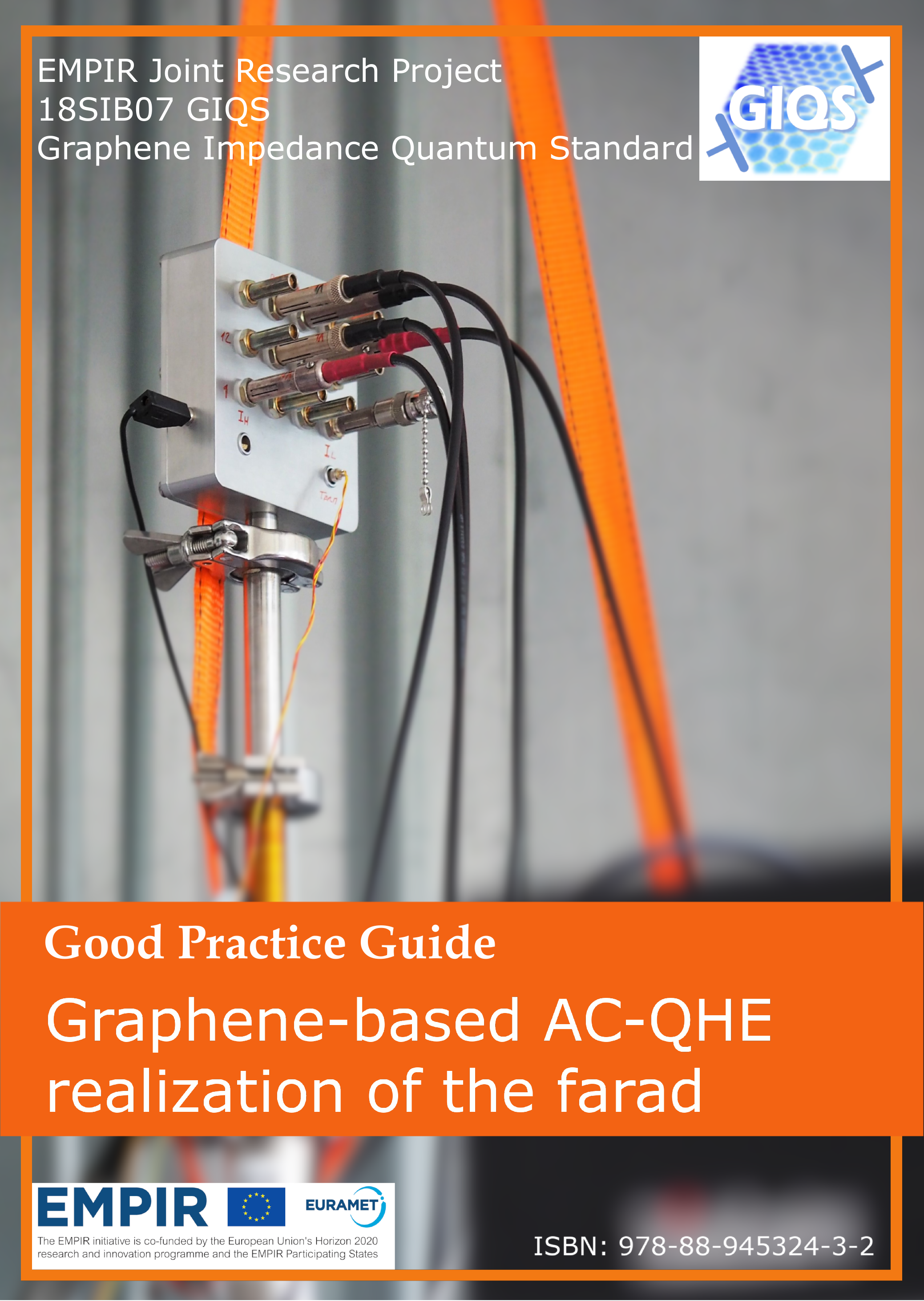}

\maketitle

\begingroup
\parindent 0pt
\parskip \baselineskip
Written and edited by:

Luca Callegaro\textsuperscript{3}, Stephan Bauer\textsuperscript{1}, Blaise Jeanneret\textsuperscript{5}, Mattias Kruskopf\textsuperscript{1}, Martina Marzano\textsuperscript{3}, 
Massimo Ortolano\textsuperscript{10,3},
Frédéric Overney\textsuperscript{5}

Contributions from: 

\begin{tabular}{llll}
1 & PTB & Physikalisch-Technische Bundesanstalt & Germany \\
2 & CMI & Cesky Metrologicky Institut & Czech Republic \\
3 & INRIM & Istituto Nazionale di Ricerca Metrologica & Italy \\
4 & LNE & Laboratoire national de métrologie et d'essais & France \\
5 & METAS & Eidgenössisches Institut für Metrologie & Switzerland \\
6 & RISE & Research Institutes of Sweden AB & Sweden \\
7 & VTT & Teknologian tutkimuskeskus Oy & Finland \\
8 & CNRS & Centre National de la Recherche Scientifique & France \\
9 & NIMT & National Institute of Metrology Thailand & Thailand \\
10 & POLITO & Politecnico di Torino & Italy \\
11 & KRISS & Korea Research Institute of Standards and Science & Republic of Korea
\end{tabular}

Cover page: Martina Marzano\textsuperscript{3} \\
ISBN: 978-88-945324-3-2 \\
May 2022 \\

\includegraphics[width=0.4\linewidth]{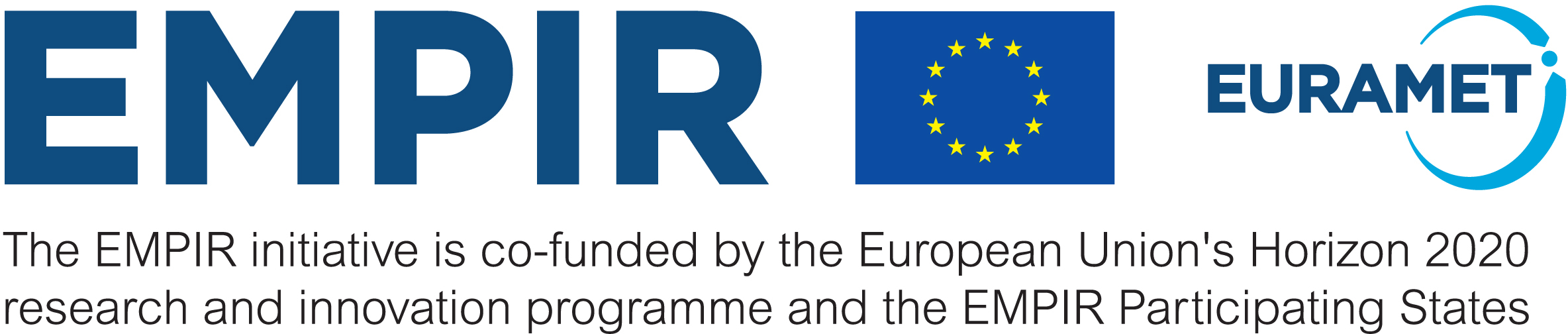}

{\footnotesize This Good Practice Guide is a deliverable of the Joint Research Project EMPIR 18SIB07 GIQS: \emph{Graphene Impedance Quantum Standard}. This project received funding from the European Metrology Programme for Innovation and Research (EMPIR) co-financed by the Participating States and from the European Unions' Horizon 2020 research and innovation programme.}
\newpage
\section*{Guide information}
\subsection*{What is this about?}
This guide provides information for the realization of the farad from the quantum Hall resistance in graphene devices by using digital impedance bridges.
\subsection*{Who is it for?} 
This guide is for researchers active in electrical metrology (national metrology institutes, calibration centers, metrology laboratories in the academy) who need to perform impedance measurements traceable to the International System of Units.
\subsection*{What is its purpose?} 
This guide provides a detailed description of how to set up a laboratory to perform measurements of artifact capacitance standards in terms of the quantized Hall resistance in graphene devices. The fabrication and characterization of graphene quantum Hall effect devices, the cryogenic environment required to achieve the quantization conditions, the digital impedance bridges and calibration procedures are reported.
\subsection*{What is the prerequired knowledge?}
This guide is for users in research and industry who have experience with and access to the advanced techniques described herein. It is targeted at researchers having experience in electrical metrology.
%

%
\clearpage
\tableofcontents
\clearpage
\listoffigures 
\listoftables
\clearpage
%
\section*{Introduction}
\emph{GIQS - Graphene Impedance Quantum Standard} \cite{GIQS} is a three-year Joint Research Project (code 18SIB07) of the European Metrology Programme for Innovation and Research (EMPIR). 

The aim of GIQS is to enable an economically efficient traceability of measurements of impedance quantities (resistance, capacitance, inductance) to the defining constants (the Planck constant and the elementary charge) of the revised International System of Units (SI). New and easier to operate measurement bridges, convenient and easier to use graphene quantum standards, cryogenic systems, and methods to combine them will be developed. 

The overall objective of GIQS is to combine novel digital impedance measurement bridges with graphene-based ac quantized Hall resistance standards in a simplified cryogenic environment, and disseminate the technology to national metrology institutes, calibration centers, research institutions and industry. Specific objectives are:
\begin{itemize}
\item To optimize graphene devices for their use in the ac regime under relaxed experimental conditions (temperature of \SI{4}{\kelvin} or higher, magnetic field less than \SI{6}{\tesla});
\item To advance digital and Josephson impedance bridges working in a wide range of impedance and frequency; 
\item To combine the graphene devices with the bridges developed, and realize a quantum capacitance standard with accuracy in the 0.1 to 0.01 ppm range; 
\item To develop a cryocooler system hosting both Josephson and quantum Hall devices as the core element of an integrated quantum resistance and impedance standard;
\item To facilitate the take up of the technology and measurement infrastructure developed in the
project by the measurement supply chain: graphene manufacturers, standards developing
organisations and end users.
\end{itemize}

\newpage

\section{The SI and the quantum Hall effect}
The International System of Units (SI) in its present form was approved by the 26th General Conference of Weights and Measures in November 2018 \cite{CGPM2018Res1} and implemented on 20 May 2019. It is described in the \emph{SI Brochure}, 9th edition \cite{SIBrochure}. The electrical units are defined by the values of the elementary charge $e = \SI{1.602176634E-19}{\coulomb}$ and the Planck constant $h = \SI{6.62607015E-34}{\joule\second}$, fixed as exact in the definition of the ampere and the kilogram.

\subsection{Units of electrical impedance}
In the SI, the units of electrical impedance are the ohm (\si{\ohm}), the farad (\si{\farad}) and the henry (\si{\henry}). They are related by 
\begin{equation}
\SI{1}{\ohm} = \SI{1}{\henry\per\second} = \SI{1}{\farad^{-1} \second}.
\end{equation}
A realization of one of the units can thus form the basis for the realisation of the other two, provided that a realization of the second or the hertz is also available.


\subsection{The quantum Hall effect}
Edwin Hall discovered the Hall effect in 1879~\cite{Hall:1879}: if a conducting slab carries an electric current $I$, and a static magnetic field $B$ is applied orthogonally to the slab surface, a voltage $V$ develops across the slab, perpendicularly to both the current direction and the magnetic field. The ratio $R_\mathrm{H} = V/I$ is called the \emph{Hall resistance}, and in the normal Hall effect its value is proportional to $B$ and dependent on slab material, thickness and temperature. 

In 1980, Klaus von Klizing discovered the \emph{quantum Hall effect}~\cite{Klitzing:1980}: in samples where a two-dimensional layer of conduction is present, for high $B$ and low temperature $T$, the Hall resistance becomes \emph{quantized}. Its value is no longer dependent on the device material or the temperature: it is a fraction $R_\mathrm{K}/i$ (where $i$ is a small integer, typically $i=2$) of a fundamental constant, the \emph{von Klitzing constant}, or \emph{quantum of resistance},  $R_\mathrm{K} = h/e^2$.
%

The samples investigated by von Klitzing were silicon devices. In the 1990s gallium arsenide (GaAs) heterostructures were developed~\cite{Piquemal:1993}, displaying a robust effect at a magnetic field in the order of \SI{10}{\tesla} and at measurement currents ($>\SI{50}{\micro\ampere}$) suited for the calibration of artifact standard resistors. The temperature required by either Si or GaAs devices are typically of the order of \SI{1.5}{\kelvin} or less.
\subsection{Realization of impedance units}
According to 9th SI Brochure, Appendix 2~\cite{SI9thBrochure_MisePratiqueAmpere},

\begin{displayquote}
\textbf{The ohm} \si{\ohm} can be realized as follows [\ldots] by using the quantum Hall effect in a manner consistent with the CCEM Guidelines and
the following value of the von Klitzing constant 
\begin{equation}
\label{eq:RKvalue}
R_\mathrm{K} = \SI{25 812.807 459 3045}{\ohm}.
\end{equation} [\ldots] 
\end{displayquote}

\begin{displayquote}
\textbf{The farad} \si{F} can be realized [\ldots] by comparing the impedance of a known resistance obtained using the quantum Hall effect and the value of the von Klitzing constant given in Eq.~\ref{eq:RKvalue}, including a quantized Hall resistance itself, to the impedance of an unknown capacitance using, for example, a quadrature bridge [\ldots]
\end{displayquote}

\begin{displayquote}
\textbf{The henry} \si{\henry} can be realized [\ldots] by comparing the impedance of an unknown inductance to the impedance of a known
capacitance with the aid of known resistances using, for example, a Maxwell-Wien bridge, where the known capacitance and resistances have been determined, for example, from the
quantum Hall effect and the value of $R_\mathrm{K}$ given in Eq.~\ref{eq:RKvalue} [\ldots]
\end{displayquote}

The impedance units can therefore be realized in terms of the quantum Hall effect, using impedance bridges. The focus of this guide is on the realization of the capacitance unit by exploiting the quantum Hall effect in graphene devices, and digital impedance bridges.

%
%
%

\section{Graphene}
\subsection{The quantum Hall effect in graphene}

The discovery of graphene (2004)~\cite{Novoselov:2004} initiated research to exploit the quantum Hall effect in this new material. It was demonstrated~\cite{Ribeiro:2015, Kruskopf:2016} that, with respect to GaAs devices, graphene QHE devices can operate at lower magnetic field (\SI{5}{T} or less), higher measurement current (up to hundreds of \si{\micro\ampere}) and higher temperature (\SI{5}{\kelvin}). These conditions allow to implement tabletop quantized Hall resistance standards using small, inexpensive dry cryocoolers~\cite{Janssen:2015,Rigosi:2019}, suitable to be continuously operated in a calibration laboratory. Research is now focusing on achieving better reproducibility of fabrication and long-term stability of devices. 

\subsection{Fabrication of graphene samples}
Here the steps for the production of graphene QHR devices are described. Section~\ref{sec:growth} is about the growth of graphene films including all involved preparation steps, Section~\ref{sec:device_fabrication} involves the lithographic processes to realize a QHR device in form of a graphene Hall bar. The described processes and measurements apply to the devices produced and investigated in the cleanroom facility and laboratories of PTB; other fabrication techniques that achieve high-quality QHR devices have been considered \cite{Ribeiro:2015,He:2018,Rigosi:2019b}.

\subsubsection{Graphene growth}
\label{sec:growth}
\begin{figure}[t]
\centering
\includegraphics[width=0.8\linewidth]{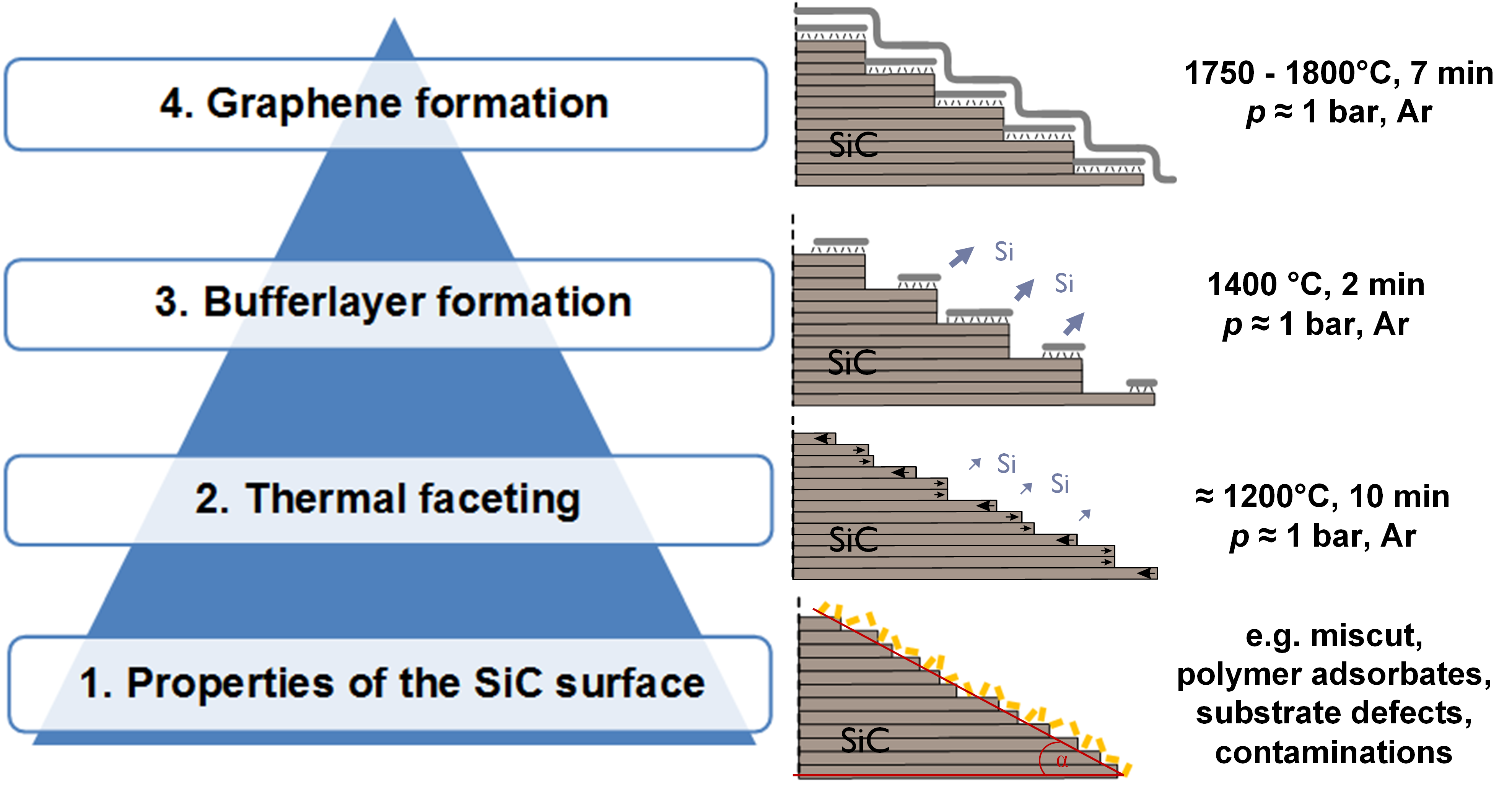}
\caption[Growth of graphene on SiC substrate]{Process steps involved in the growth of graphene on SiC substrates - from the bottom to the top.}
\label{fig:PTB_process1}
\end{figure}
\begin{figure}[tb]
\centering
\includegraphics[width=0.7\linewidth]{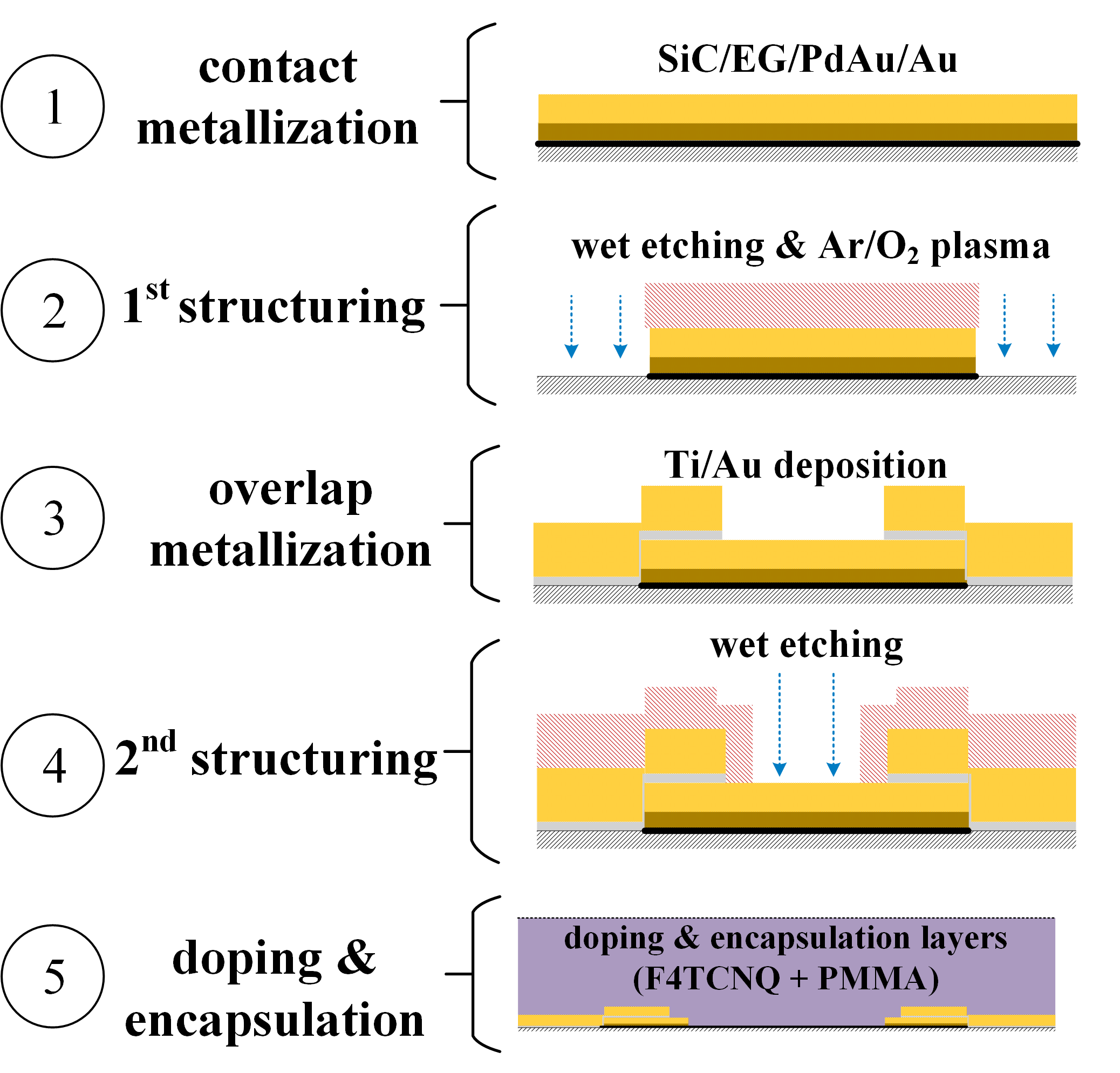}
\caption[Fabrication of graphene Hall bars]{Photolitography steps for the fabrication of graphene Hall bars.}
\label{fig:PTB_process2}
\end{figure}
\begin{figure}[tb]
\centering
\includegraphics[width=0.9\linewidth]{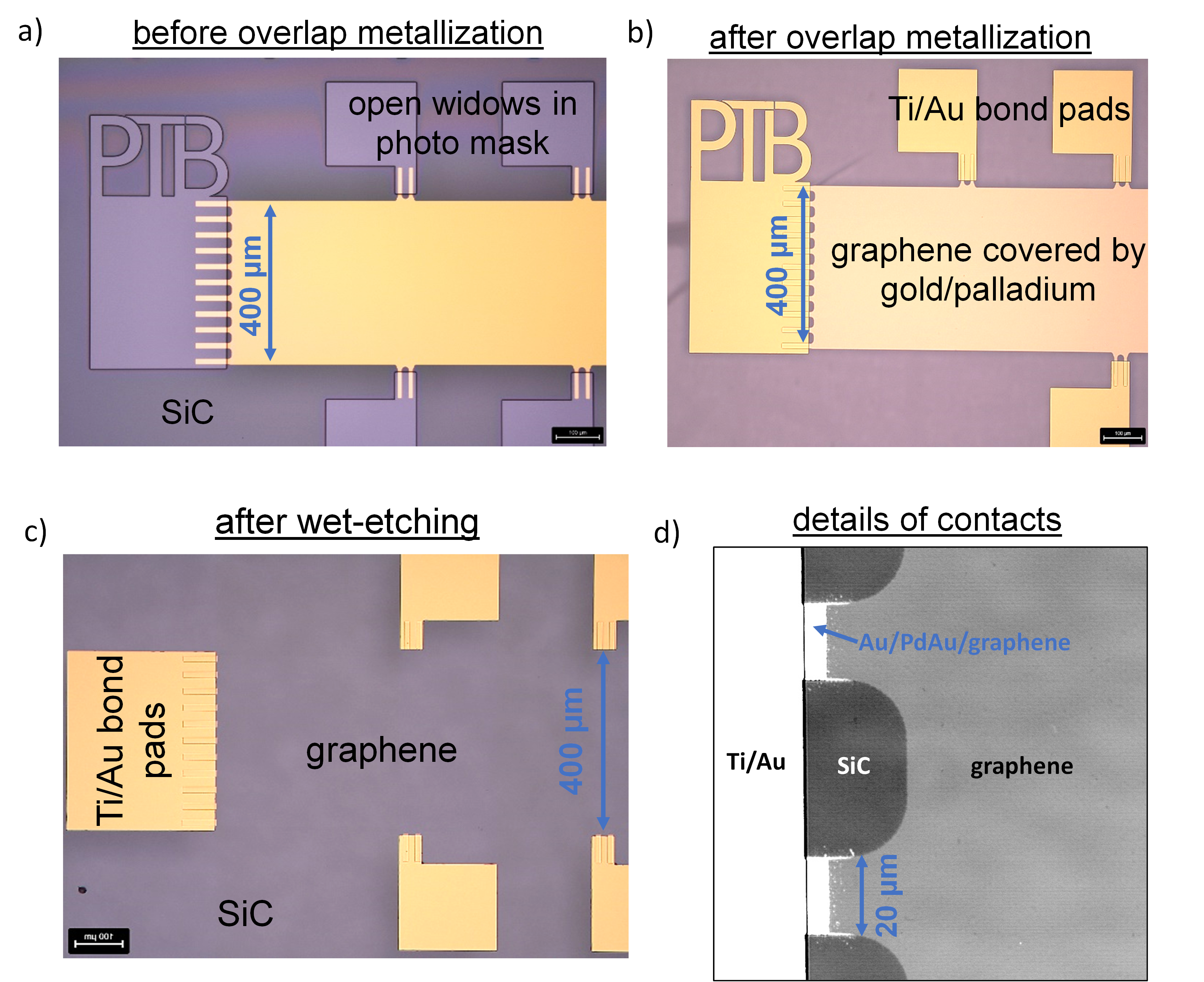}
\caption[Microscope images of graphene Hall bars during processing]{Microscope images of graphene QHR Hall bars at different steps of the device processing. (a) Structured Hall bar and the developed photo mask befrore the deposition of Ti/Au. (b) after Ti/Au deposition,with the graphene still being covered and protected by the Au/PdAu layer. (c) After selective removal of the metal cover of the graphene film by wet-etching. (d) Detail image of the split-contact design, showing$\approx \SI{20}{\micro\metre}$ wide graphene fingers that are separated from each other by$\approx \SI{40}{\micro\metre}$.}
\label{fig:PTB_process3}
\end{figure}
The processes shown in Figure~\ref{fig:PTB_process1} describe, from bottom to top, the four major steps to consider for the growth of high quality monolayer graphene films. 

Realizing the appropriate starting properties of the SiC before the thermal annealing is initiated in the growth chamber is a very crucial step. Suggested substrates are prime-grade (low defect density) 4H and 6H-SiC wafers with a CMP prepared Si-face with a miscut angle $\leq \SI{0.15}{\degree}$. Semi-insulating or n-type wafers may be used since both types of wafers are insulating at temperatures around \SI{4}{\kelvin}. The substrate should be free of metal or organic contaminations which can originate from the processes involved during the dicing procedure of the wafer. Dicing tapes may be used to protect the Si-face of the wafer, but intensive cleaning must be applied before processing is continued. Usually, a batch of diced substrate pieces is kept in a beaker with acetone for at least \SI{12}{\hour}. On the day the diced substrate pieces are introduced into the growth chamber, they are cleaned in ULSI grade acetone and ULSI isopropanol for \SI{10}{\minute} and \SI{15}{\minute}, respectively, and spin-dried afterward. The substrates then follow a polymer treatment for Polymer-Assisted Sublimation Growth (PASG) by spin-on deposition at 6000 rpm. The suitable volume ratio of the polymer solution is \SI{2.3}{\micro\litre} concentrated AZ5214E photoresist solved in \SI{1}{\micro\litre} isopropanol. 

The so prepared substrates are then introduced into the growth chamber to initiate the growth process following the evacuation of the growth chamber. The base pressure to be reached before the annealing process is initiated is \SI{E-6}{\milli\bar}, after which the chamber and samples are heated in vacuum\footnote{This step is for additional cleaning and decomposition of the polymer adsorbate into amorphous carbon.} for \SI{30}{\minute} at \SI{900}{\celsius}. The thermal faceting of the substrate is initiated at \SI{1200}{\celsius} for \SI{10}{\minute}. Then the temperature is raised to \SI{1400}{\celsius}, where Si sublimation of the substrate starts and the deposited amorphous carbon converts into a high density of buffer layer domains. The high density of buffer layer islands prevents the formation of high step edges, so-called giant steps, during the subsequent graphene growth at a temperature of \SI{1750}{\celsius} and \SI{1800}{\celsius} for \SI{7}{\minute} in total.
\subsubsection{Device fabrication}
\label{sec:device_fabrication}
For device fabrication, a UV-photolithography process is used, which compared to e-beam lithography has the advantage of being much faster. The process principle originated at NIST~\cite{Yang:2015}, and was further improved at PTB with respect to different wet- and dry-etching processes, metal compositions, and photoresists during the GIQS project.

In the first step, the graphene layer, indicated by the thick black line in Figure~\ref{fig:PTB_process2}, is covered by a PdAu/Au layer which is then structured by wet-etching using potassium iodide solution and O$_2$ plasma in the second step. In step 3, the structured Hall bar, which is at this point still covered by PdAu/Au, is partially overlapped by Ti/Au pads in the regions that later become the electrical contacts. In step 4, the metal covering the Hall bar is selectively removed by wet etching using a photomask. Microscope images from steps 2, 3, 4, and graphene/metal contacts are given in Figure~\ref{fig:PTB_process3}. For chemical adjustment of the charge carrier density, the device is covered by doping and encapsulation layers using F4TCNQ molecules as dopant and PMMA resist for encapsulation \cite{He:2018,He:2019}. Subsequently, the devices are wire bonded onto a chip carrier.

The Hall bar geometry for devices used for standard dc measurements (Figure~\ref{fig:PTB_devicescheme}) has eight contacts of which six are Hall contacts, and two are the source and drain contacts. For the purpose of ac operation, the number of Hall contacts was reduced to the minimum number of six contacts that are required for the triple-series connection \cite{Chae:2022}. Both the distance between two neighboring contacts and the width of the graphene channel are \SI{400}{\micro\metre}. Electrical contacting is realized by split-contacts with eleven individual fingers in case of the source/drain contacts and two fingers in the Hall contacts as it can be understood from \ref{fig:PTB_process3}(a). The split contact helps to minimize the contact resistance of the device, and typical values of the order of a few \si{\milli\ohm} or even on the \si{\micro\ohm} level can be realized \cite{Kruskopf:2019b}. 
\subsection{Characterization}
\begin{figure}[tb]
\centering
\includegraphics[width=0.5\linewidth]{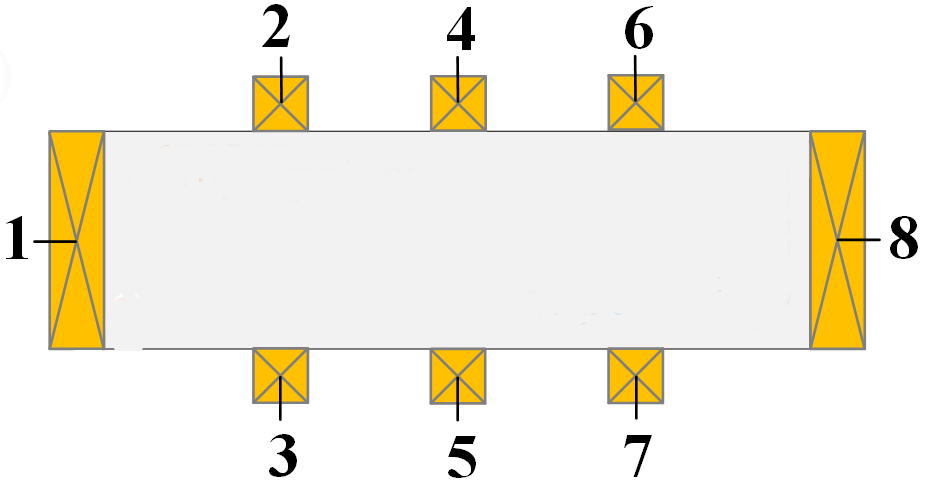}
\caption[Electrical contact labels]{Device schematic of the electrical contact labels and equi-potentials in the quantum Hall regime.}
\label{fig:PTB_devicescheme}
\end{figure}
\begin{table}[tb]
\centering
\begin{tabular}{cc|cccccccc}
\toprule
$T$ & cooldown & $R_{2.4}$ & $R_{4,6}$ & $R_{3,5}$ & $R_{5,7}$ & $R_{1,8}$ & $R_{2,3}$ & $R_{4,5}$ & $R_{6,7}$\\
\si{\kelvin} & \si{\minute} & \si{\kilo\ohm} & \si{\kilo\ohm} & \si{\kilo\ohm} & \si{\kilo\ohm} & \si{\kilo\ohm} & \si{\kilo\ohm} & \si{\kilo\ohm} & \si{\kilo\ohm} \\
\midrule
294 & 0 & 6.2 & 5.8 & 6.2 & 5.9 & 25.5 & -0.14 & -0.13 &-0.03 \\
4.2 & 30 & 3.9 & 3.8 & 4.0 & 3.9 & 16.5 & -0.28 & -0.24 & -0.15 \\
\bottomrule
\end{tabular}
\caption[Typical four-terminal resistances of graphene QHE devices]{Typical four-terminal resistances of graphene QHE devices measured at RT and at \SI{4.2}{\kelvin} and $B = \SI{0}{\tesla}$ with fixed current terminals using pin 1 and 8.}
\label{tab:PTB_4res}
\end{table}
\begin{figure}[p]
\centering
   \begin{subfigure}[a]{0.7\textwidth}
         \centering
         \includegraphics[width=\textwidth]{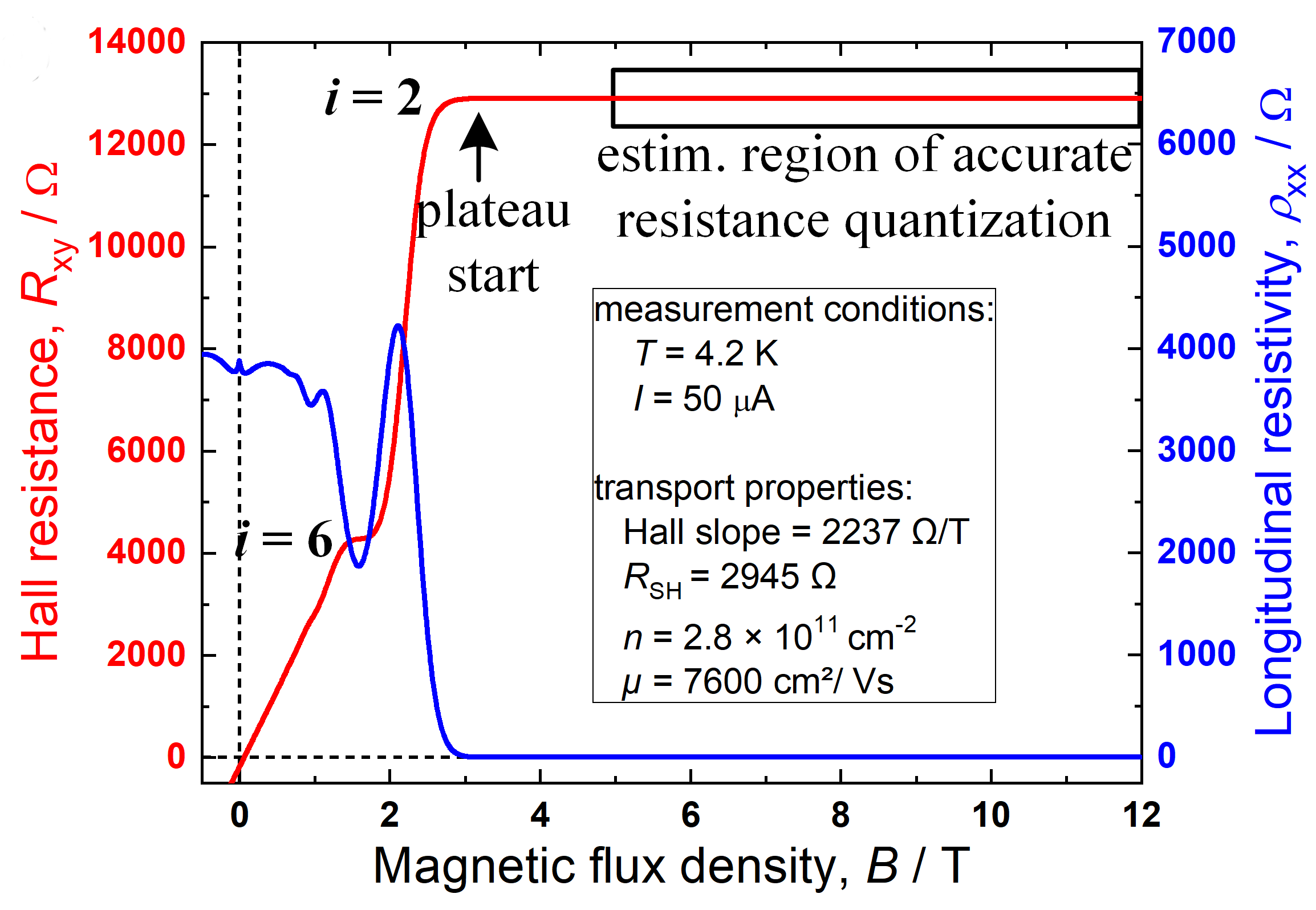}
         \caption{}
         \label{fig:PTB_magnetotransport_a}
     \end{subfigure}
   \begin{subfigure}[b]{0.7\textwidth}
         \centering
         \includegraphics[width=\textwidth]{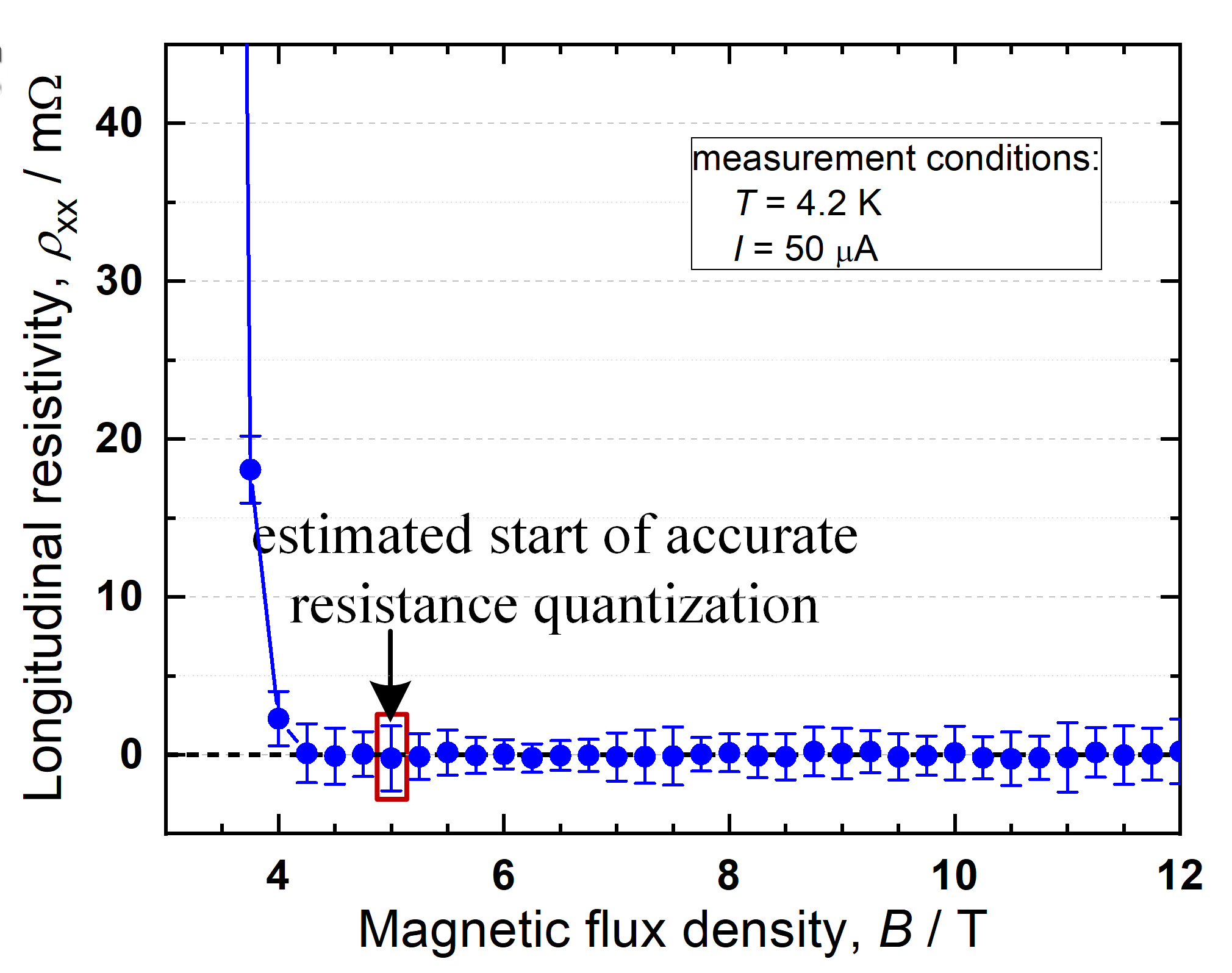}
         \caption{}
         \label{fig:PTB_magnetotransport_b}
     \end{subfigure}
\caption[Magnetotransport measurements of a graphene QHE device]{Typical magnetotransport measurements of a graphene-based QHE device. (\subref{fig:PTB_magnetotransport_a}) The start of the resistance plateau (black arrow) and the expectable range of $B$-field values where accurate resistance quantization with $R_\mathrm{xy} = R_\mathrm{K}/2$ is achieved (black square box) are marked accordingly. The electrical transport properties in the inset are typical for devices with accurate resistance quantization starting around $B = \SI{5}{\tesla}$ at the given conditions. (\subref{fig:PTB_magnetotransport_b}) Longitudinal resistances determined by the current-reversal measurement technique at fixed $B$-field values using a precision current source and a nano-voltmeter. For separation of $\Delta B = \SI{0.25}{\tesla}$ and the given measurement conditions, the 4th $\rho_{xx}$ value that is zero within the measurement uncertainty typically gives the margins for accurate resistance quantization. The error bar indicates the type A uncertainty ($k = 1$).}
\label{fig:PTB_magnetotransport}
\end{figure}
\begin{table}[tb]
\centering
\begin{tabular}{cccc}
\toprule
contact/pin & current & voltage & resistance \\
 & contacts & contacts & \si{\ohm}\\
\midrule
1           &
1 \& 8           &
1 \& 2           &
1.2\\           
2           &
2 \& 7           &
2 \& 4           &
1.5\\
4           &
4 \& 5           &
4 \& 6           &
1.6\\           
6           &
6 \& 3            &
6 \& 8           &
1.5\\
8           &
8 \& 1           &
8 \& 7           &
1.2\\
7           &
7 \& 2           &
7 \& 5           &
1.5\\
5           &
5 \& 4           &
5 \& 3           &
1.6\\         
3           &
3 \& 6          &
3 \& 1          &
1.5\\
\bottomrule
\end{tabular}
\caption[Three-terminal contact resistance measurements]{Measurement procedure for three-terminal contact resistance measurements within the well-quantized resistance plateau. The resistance values represent a practical example performed at \SI{4.2}{\kelvin} and $B = \SI{6}{\tesla}$ in a system with non-coaxial cables.}
\label{tab:PTB_contacts}
\end{table}
The electrical characterization of graphene devices follows several steps, from pre-characterization to high accuracy measurements. The framework for precision quantum Hall measurements are the existing guidelines of quantum Hall resistance metrology for GaAs QHE devices~\cite{Delahaye:2003}. However, since the handling and behaviour of graphene QHE devices are different from those of GaAs devices, the steps involved in the electrical characterization including the precision measurements require specific adjustments of the measurement protocol.

The following measurement protocol for graphene QHR devices starts with introducing the measurement methods that are important to characterize the device quality prior to high accuracy quantum Hall measurements. These preliminary investigations comprise, e.g., measurements of the sheet resistance(s), the overall shape of the Hall curve over a wide range of magnetic flux densities and the contact resistances. In Sections~\ref{sec:resistances}, \ref{sec:magnetotransport}, \ref{sec:contacts}. In Sections~\ref{sec:dchall}, \ref{sec:longitudinal}, \ref{sec:validation} a dc resistance bridge is used to evaluate the so-called $s$-parameter and to validate the relevant device characteristics with the highest possible precision. 

The measurement protocol is explained with the help of a practical example. The schematic in Figure~\ref{fig:PTB_devicescheme} shows the labeling of the electrical contacts and the typical configuration of the electrical potentials in the quantum Hall regime.

\subsubsection{Room-temperature resistances and cooldown procedure}
\label{sec:resistances}
Before the device (mounted on the chip carrier) is cooled down in the cryostat, it is attached to the probe stick at room temperature for simple room temperature characterization to identify open contacts or inhomogeneities in its electrical properties. 

Graphene quantum Hall devices with an electron charge carrier density on the order of \SIrange{1e-11}{2e-11}{\centi\meter\tothe{-2}} typically have a room temperature sheet resistance around \SIrange{6}{7}{\kilo\ohm}. Table~\ref{tab:PTB_4res} shows electrical resistances measured at room temperature (RT) as well as after the cooldown at \SI{4.2}{\kelvin}. All resistances $R_{i,j} =  U_{i,j}/I_{1,8}$ are four-terminal measurements with fixed current ports at pin 1 and pin 8 while the voltage ports are changed as given by the indices $i$ and $j$. During the cooldown procedure, which is typically performed within \SI{30}{\minute}, all contacts are shorted to each other as well as to the cryostat. A device is considered \emph{good} if the sheet resistances ($R_{2,4}, R_{4,6}, R_{3,5}, R_{5,7}$) are similar within a few hundred ohms, and if the zero-field Hall resistances ($R_{2,3}, R_{4,5}, R_{6,7}$) are below \SI{500}{\ohm}. Note, that these figures are no sharp limits, but they rather reflect the margins (typical variation) of the electrical property values of high-quality graphene devices. However, devices with anomalies in the Hall and sheet resistances at zero magnetic ($B$) field indicated the presence of defect contacts or pronounced material inhomogeneities and thus should not be used for calibration or other purposes requiring ultimate accuracy.
\subsubsection{Magnetotransport properties, Hall measurements}
\label{sec:magnetotransport}
For the characterization of the magnetotransport properties a sweep of the magnetic flux density is performed while measuring the longitudinal resistivity $\rho_{xx}$ ($= R_{3,7} / 2$) and the Hall resistance $R_{xy}$ ($= R_{4,5}$).

Figure~\ref{fig:PTB_magnetotransport} shows results of a typical magnetotransport measurement on a graphene-based QHE device for magnetic flux densities up to \SI{12}{\tesla}, performed with a precision dc current source and a dc nano-voltmeter (alternatively, a lock-in amplifier can be used).

Depending on the charge carrier density, the start of the $i = 2$ plateau with $R_{xy} = R_\mathrm{K}/2$ is found at magnetic flux densities as low as a few \si{\tesla}. While the device in Figure~\ref{fig:PTB_magnetotransport_a} shows a quantum Hall plateau starting at around \SI{3}{\tesla} (depending on the measurement temperature), the first point of accurate resistance quantization on the level of \SI{1}{\nano\ohm\per\ohm} is found at a B-field that is typically \SIrange{1}{2}{\tesla} higher than the beginning of the plateau, as marked by the black square box. 

In addition to the shape and the onset of the resistance plateau, the Hall slope $s_\mathrm{Hall}$, the sheet resistance $R_\mathrm{SH}$ as well as the charge carrier density $n = 1/(s_\mathrm{Hall} \cdot e)$ and the charge carrier mobility $\mu = 1/(n \cdot R_\mathrm{SH} \cdot e)$ are important properties that must be documented to judge the quality of the device in terms of QHR application. Typical electron densities of devices that start to be accurately quantized between \SI{4}{\tesla} and \SI{6}{\tesla} are $n =\SI{1E11}{cm^{-2}}$ to $n = \SI{3E11}{\cm^{-2}}$  with electron mobilities $\mu \geq \SI{5000}{\cm^2 \per (\volt \second)}$.

A suitable way of pre-characterizing the start of the resistance plateau with higher precision is by measuring the longitudinal resistance with a precision current source and a nano-voltmeter at fixed $B$-field points. To suppress thermal voltages due to temperature gradients in cables and connectors the current reversal measurement technique is applied. Representative results that were obtained using such a relatively simple measurement setup (with a measurement uncertainty on the level $\leq \SI{5}{\milli\ohm}$ are shown in Figure~\ref{fig:PTB_magnetotransport_b}. In this way, the $B$-field range of “accurate resistance quantization” can be estimated and prominent anomalies in the longitudinal resistance can be identified. 

With a $\Delta B = \SI{0.25}{\tesla}$ between the individual $\rho_{xx}$ measurements and with $I = \SI{50}{\micro\ampere}$ at $T = \SI{4.2}{\kelvin}$, “accurate quantization” typically starts not before the 4th $\rho_{xx}$ value drops to zero within the measurement uncertainty. However, even if no anomalies can be identified in the $\rho_{xx}$ measurements, the estimation of the starting point of accurate quantization is just a preparation step to select an appropriate range of magnetic flux densities to be applied in the subsequent parts of the measurement protocol. 
\subsubsection{Contact resistance measurements}
\label{sec:contacts}
For detailed contact resistance investigations, the same measurement equipment as in \ref{sec:magnetotransport} is used in combination with in the current reversal measurement technique in a three-terminal configuration. The measurements are performed at a fixed $B$-field values within the margins of the resistance plateau where the longitudinal resistance is zero within the measurement (see Figure~\ref{fig:PTB_magnetotransport_b}). 
The measurements in Table~\ref{tab:PTB_contacts} represent typical values that are obtained in a cryostat with cable resistances being of the order of \SI{1}{\ohm}. Since the three-terminal measurements also include cable resistances and any other resistances along the current-carrying line that is part of the voltage signal path, the true contact resistances are always lower than the measured values. After considering the included resistances of cables and connectors, the remaining contact resistance of a typical device fabricated by the techniques described in Section~\ref{sec:device_fabrication} is below \SI{1}{\ohm}. However, also higher contact resistances below \SI{10}{\ohm} are acceptable \cite{Delahaye:2003}.
\begin{figure}[p]
\centering
   \begin{subfigure}[a]{0.7\textwidth}
         \centering
         \includegraphics[width=\textwidth]{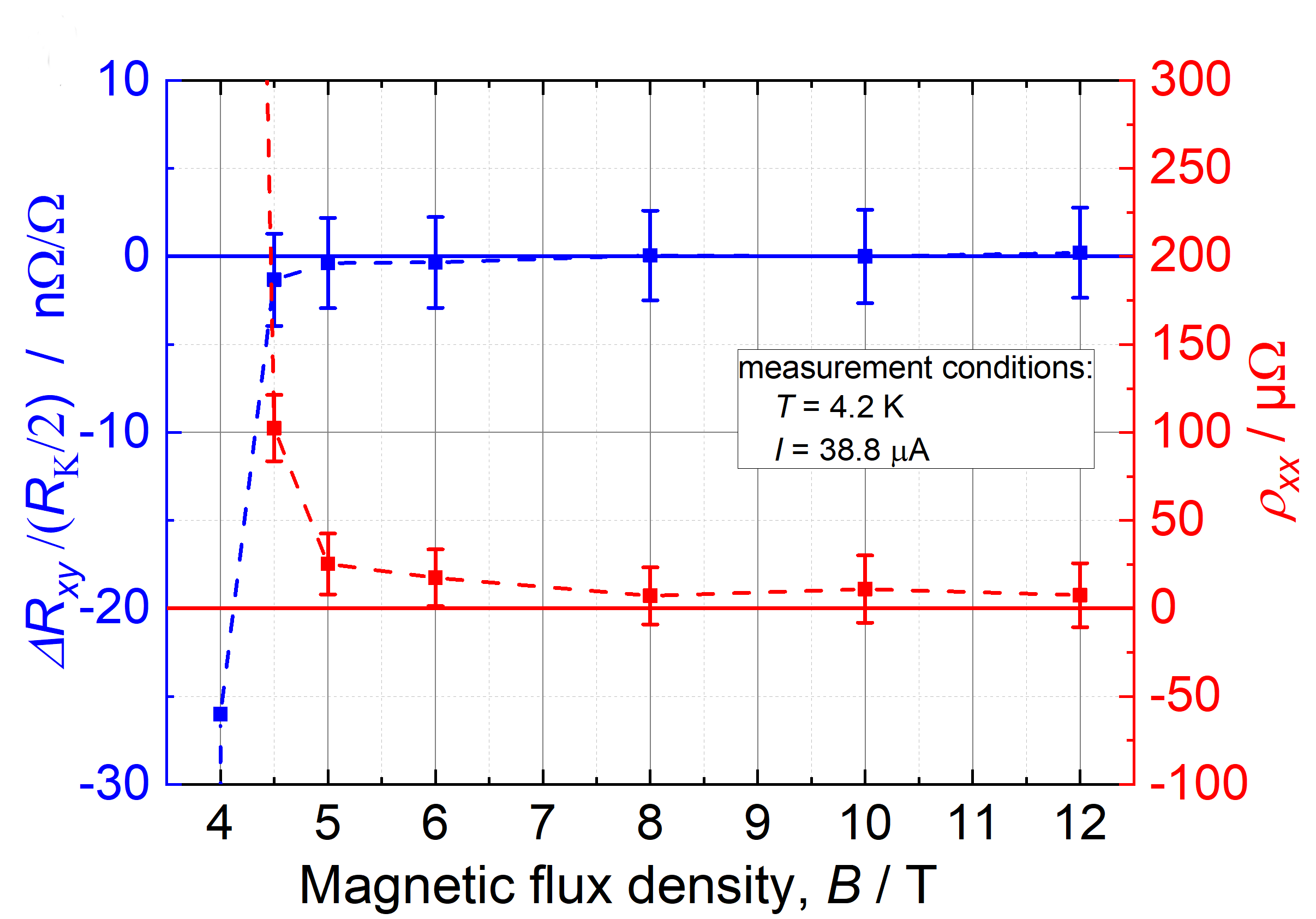}
         \caption{}
         \label{fig:PTB_precision_a}
     \end{subfigure}
   \begin{subfigure}[b]{0.7\textwidth}
         \centering
         \includegraphics[width=\textwidth]{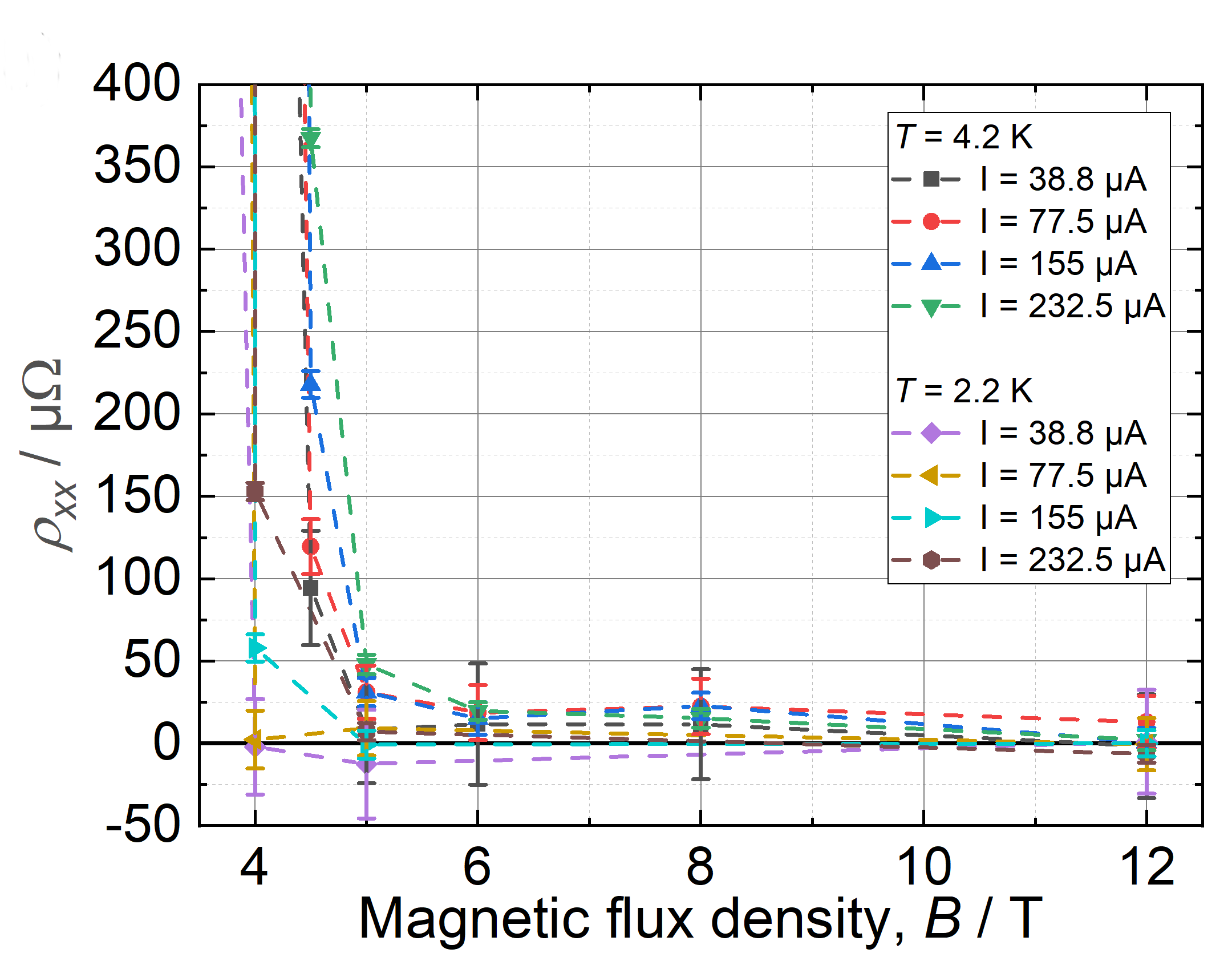}
         \caption{}
         \label{fig:PTB_precision_b}
     \end{subfigure}
\centering
\caption[Precision measurements of Hall and longitudinal resistance]{Precision measurements of the Hall and the longitudinal resistance. (\subref{fig:PTB_precision_a}) In the regions with the longitudinal resistivity being below $\rho_{xx} < \SI{50}{\micro\ohm}$, the deviation of the Hall resistance from $R_\mathrm{K}/2$ is consistent with zero within the measurement uncertainty. (\subref{fig:PTB_precision_b}) Measurements at higher currents and different temperatures demonstrate the robustness of the QHR devices. As a result of the robustness, temperature and current variations are of limited use for the determination of the s-parameter within the resistance plateau. All uncertainty figures correspond to combined expanded uncertainties ($k=2$).}
\label{fig:PTB_precision}
\end{figure}
\begin{table}[tb]
\centering
\begin{tabular}{cccc}
\toprule
$B$ & $\rho_{xx}$ & $\Delta R_{xy}$ & $s$-parameter \\
\si{\tesla} & \si{\micro\ohm} & \si{\micro\ohm} & \\
\midrule
\num{3.5}       &
\num{168499 \pm 306} &
\num{-51608 \pm 445} &
\num{-0.31 \pm 0.003} \\
\num{4.0}       &
\num{1602.3 \pm 17.5}   &
\num{-335.8 \pm 33.7}   &
\num{-0.21 \pm 0.02} \\
\num{4.5}       &
\num{102.5 \pm 18.8}    &
\num{-17.1 \pm 33.8}    &
\num{-0.17 \pm 0.30}    \\
\num{5.0}       &
\num{25.3 \pm 17.3}     &
\num{-4.8 \pm 33.1}     &
\num{-0.19 \pm 1.31}    \\
\num{6.0} &
\num{17.4 \pm 16.1}     &
\num{-4.4 \pm 33.3}     &
\num{-0.25 \pm 1.9}     \\
\bottomrule
\end{tabular}
\caption[$s$-parameter determination]{$s$-parameter values determined at the edge of the quantum Hall plateau. $s$-parameter values that were determined at the outer edge of the plateau may be used as an estimate for s-parameter values within the plateau. Inside the resistance plateau with $\rho_{xx}$ values below \SI{100}{\micro\ohm} (in this example at $B = \SI{5}{\tesla}$), the uncertainty of the $s$-parameter becomes too large. Uncertainties figures are combined expanded uncertainties ($k=2$).}
\label{tab:sparameter}
\end{table}%
\begin{figure}[tb]
\centering
\includegraphics[width=\linewidth]{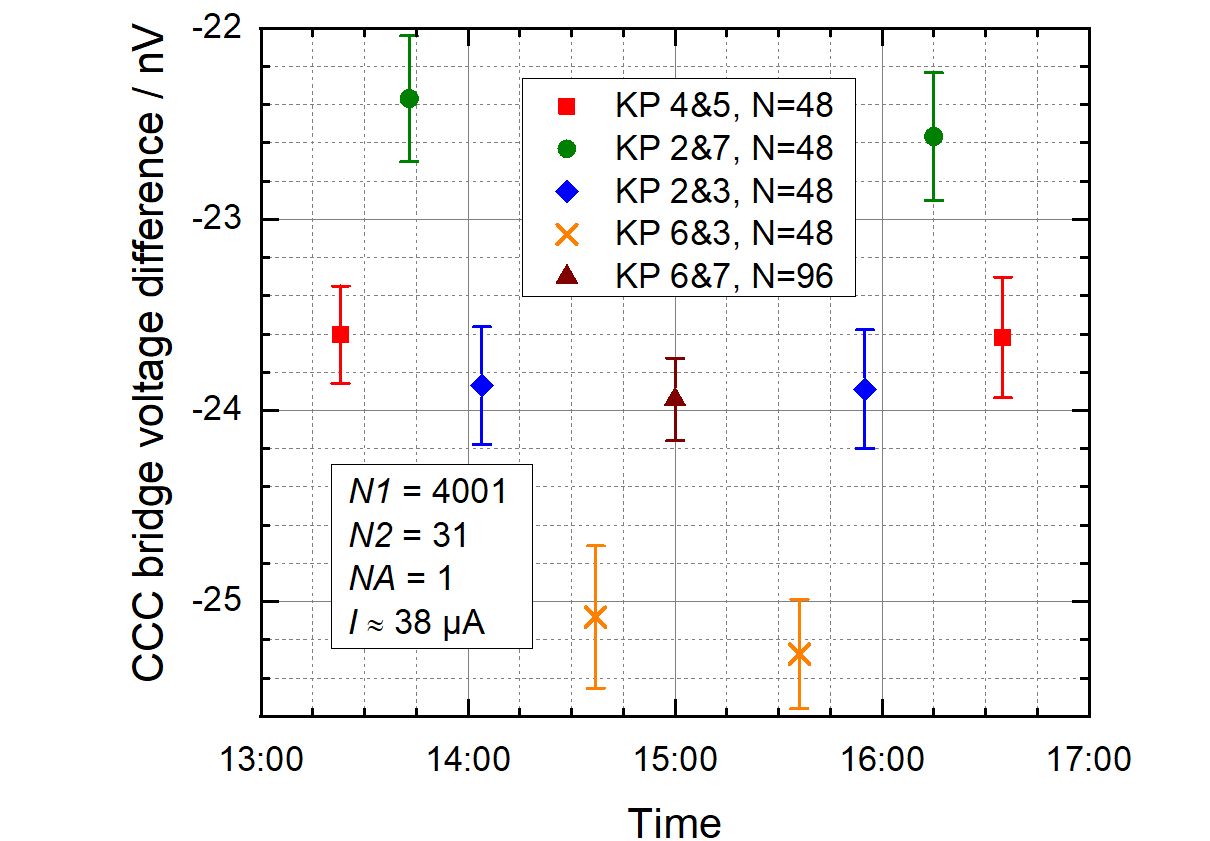}
\caption[Final device validation]{An example of the final precision device validation prior to calibration at a fixed $B=\SI{6}{\tesla}$ where the deviation from $R_\mathrm{K}/2$ is expected to be below \SI{1}{\nano\ohm\per\ohm}. In this final characterization step of the device, a series of CCC measurements is performed that involves all Hall contacts with the primary purpose to identify instabilities over time as well as inhomogeneities in the device properties. The quantity $N$ describes the number of measurement cycles in the CCC measurement. The uncertainty figures are type A expanded measurement uncertainties ($k = 2$).}
\label{fig:PTB_validation}
\end{figure}
\subsubsection{Precision dc quantum Hall measurements}
\label{sec:dchall}
While any of the previous steps apply relatively simple measurement equipment, a device must finally be checked with high accuracy before it is used for calibration purposes. Suitable measurement systems are CCC (Cryogenic Current Comparator) or DCC (Direct Current Comparator) resistance bridges that allow determining the $R_\mathrm{K}/2$ value with a type A uncertainty of less than \SI{20}{\micro\ohm} at a current of \SI{50}{\micro\ampere}. These systems can also be used to determine the longitudinal resistivity $\rho_{xx}$ from the difference of two Hall resistances measured at orthogonally and diagonally aligned contact pairs.%
\subsubsection{Evaluation of the longitudinal resistivity and the $s$-parameter}
\label{sec:longitudinal}
The main quantization criterium is the vanishing of the longitudinal resistivity $\rho_{xx}$ such that the admixing of the longitudinal contribution becomes negligible. Since in practice the longitudinal resistivity is often low, but non-zero, $\rho_{xx}$ values below \SI{50}{\micro\ohm} are still acceptable if the $s$-parameter can be evaluated and if it is sufficiently small.

The $s$-parameter $s = \Delta R_\mathrm{xy} / \rho_{xx}$ describes the dependency between $R_\mathrm{xy}$ and $\rho_{xx}$, where $\Delta R_\mathrm{xy} = R_\mathrm{xy}-(R_\mathrm{K}/2))$ is the deviation of the measured Hall resistance. 


In the case that $s \leq \pm 0.5$ or even $s \leq \pm 1$, the acceptable upper limit for $\rho_{xx}$ is $\rho_{xx} \leq \SI{25.8}{\micro\ohm}$ or $\rho_{xx} \leq \SI{12.9}{\micro\ohm}$ respectively.  Typically, graphene QHR devices can be operated under conditions where $s$ and $\rho_{xx}$ are within these limits. If a device is operated outside these limits, there is still the possibility to correct for the longitudinal contribution in a calculative way. Ideally, $s$ is characterized at the same $B$-field value at which the device is operated later. This can be done by varying the temperature or current until a significant increase in the longitudinal resistivity and related changes in $\Delta R_\mathrm{xy}$ are observed. However, due to the robustness of the QHE in graphene devices as well as due to parameter limits of individual cryostats and measurement systems, the $s$-parameter may alternatively be determined at the outer edge of the accurately quantized resistance plateau where $\rho_{xx}$ increases to a few \SI{100}{\micro\ohm}. When operating the device in this regime, one can determine the $s$-parameter with relatively low uncertainty figures by measuring $\Delta R_\mathrm{xy}$ and $\rho_{xx}$ while varying the temperature, current or magnetic field. An example of systems that allow temperature variations in only small margins are compact table-top cryostats that are being adopted by many NMI’s \cite{Janssen:2015, Rigosi:2019}.
In Table~\ref{tab:sparameter}, the $s$-parameter and the related quantities $\Delta R_\mathrm{xy}$, $\rho_{xx}$, were determined by varying the magnetic field. Since within the resistance plateau with $\rho_{xx}$ values below \SI{100}{\micro\ohm} the uncertainty of the $s$-parameter becomes too large, reliable $s$-parameter values are only obtained at the edge of the plateau. In the practical example, all evaluated s values are smaller than $s = \pm 0.5$. As a result, the upper limit for an acceptable longitudinal resistivity is $\rho_{xx} = \SI{25.8}{\micro\ohm}$ allow for precision resistance quantization with a deviation from $R_\mathrm{K}/2$ below or equal to $\SI{1}{\nano\ohm\per\ohm}$. As a result of the measured $\rho_{xx}$ values in Figure~\ref{fig:PTB_precision}, the lowest magnetic flux density point where the device can be expected to be sufficiently well-quantized is around $B = \SI{6} {\tesla}$.
\subsubsection{Final precision device validation}
\label{sec:validation}
Figure~\ref{fig:PTB_validation} represents the final check before the device can be used for calibrations. This systematic series of CCC measurements involves all Hall contacts of the device. It is performed at the same fixed $B$-field that is intended to be later used for the calibration procedure in which the QHR will be the reference. The suitable $B$-field was selected where the deviation from $R_\mathrm{K}/2$ is expected to be on the level of \SI{1}{\nano\ohm\per\ohm}, as described in Section~\ref{sec:longitudinal}. The main purpose of this measurement is to identify instabilities over time as well as inhomogeneities in the device properties for different combinations of contacts.

Before high accuracy measurements are performed, a pre-measurement is started for at least \SI{1.5}{\hour} to ensure that the reference resistor reaches a stable temperature. Then a series of eight time-symmetrically arranged Hall measurements at five pairs of Hall contacts (see contact labels in Figure~\ref{fig:PTB_devicescheme}) are applied in the following order: $1. \rightarrow 4 \& 5$, $2. \rightarrow 2 \& 7$, $3. \rightarrow 2 \& 3$, $4. \rightarrow 6 \& 3$, $5. \rightarrow 6 \& 7$, $7. \rightarrow 2 \& 7$, $8. \rightarrow 4 \& 5$.

While the contact pairs $2\& 7$ and $6\& 3$ are each diagonally aligned and are thus expected to display a significant longitudinal resistance component, the remaining contact pairs $4\& 5$, $2\& 3$ and $6\& 7$ are orthogonally aligned Hall contacts. Therefore, in the case of a device where all areas of the device are equally quantized, the Hall resistances and corresponding bridge voltages at the Hall contact pairs $4 \& 5$, $2 \& 3$, and $6 \& 7$ should be the same within the expanded uncertainties. Since the remaining contact pairs $2 \& 7$ and $6 \& 3$ have a longitudinal component across the full accessible length of the device, they should deviate from the previous three pairs according to their longitudinal resistance component. The results of the practical example, plotted in Figure~\ref{fig:PTB_validation}, represent a typical pattern of such a measurement series. For a known reference resistor value, winding ratio, and compensation network configuration, the value of the second resistor (device under test) can be determined from the bridge voltage difference \cite{Goetz:2009}. 

In the case of instabilities in the QHR device properties or in the reference resistor, the measurement results can be asymmetrically distributed, have different noise figures and may not be reproducible over time within the expanded uncertainties. To be able to identify instabilities caused by the reference resistor, it is recommended to continuously record the ambient pressure and resistor temperature during the measurement, primarily if dependencies are known or suspected. The final dc calibration procedure in which the QHR is used as a reference is typically realized by using the center Hall contact pair $4 \& 5$ after the device passes the complete characterization procedure discussed in this document.
\clearpage
\section{Cryogenic environment}
\subsection{Temperature and field requirements}

Two of the major advantages of graphene based QHE devices is that the requirements concerning temperature and magnetic field can be reduced compared to devices based on GaAs. The latter need typical magnetic field strengths of approx. \SI{8}{\tesla} to \SI{10}{\tesla}.  In contrast to this, graphene-based devices can be tuned to have good quantisation starting at a magnetic field strength of about \SI{3}{\tesla}. Even though lower fields are possible there is no big advantage to further reduce this operating range. On the one hand this is achieved with lowering the carrier density of the device into regions where quantization starts to get unstable. On the other hand, superconducting magnets will not change significantly in price and handling under a field of \SI{7}{\tesla}.
The temperature requirements are also relaxed if a graphene based QHE devices is used. There is no urgent need to operate the device at temperature below \SI{4.2}{\kelvin} (boiling point of Helium). Although a GaAs based QHE device can be operated at \SI{4.2}{\kelvin} \cite{Kucera:2019}, the range of the magnetic field where the device shows a quantized Hall resistance is significantly reduced. Lowering the temperature below \SI{4.2}{\kelvin} will cause some more effort (Lambda cooler, \SI{1}{\kelvin} pot, VTI,\ldots) but will reduce the thermal noise (Johnson noise) of the QHR.

\subsection{Shielding and coaxiality}
In the DC regime the quantum Hall resistance is defined as a four-terminal (4T) resistance~\cite[Sec.~2.1]{Callegaro:2013}. If the 4T definition is applied (no current is drawn from the voltage terminals by the measurement setup) the cable errors are due to the wiring parasitic conductances, which can be made negligible with adequate isolation.

In the AC regime a proper \emph{impedance definition} becomes essential. Most accurate impedance definition rely on the \emph{coaxial terminal-pair} concept  \cite[Sec.~2.2.1]{Callegaro:2013}. The QHE device must be shielded, each terminal becomes a coaxial connection to the outside, and the measurement circuit must take care of the coaxiality condition (that is, in each coaxial pair the electrical currents in the inner and outer conductor are equal and opposite). 

The coaxiality condition requires that all coaxial lines must be completely isolated from the cryostat, and be terminated by isolated coaxial connectors. The shields (outer connector) of all coaxial lines connecting the QHR, should be joint at a single point at or close to the sample holder. Such configuration is not standard (typical commercially-available coaxial lines are unisolated RF ones) and may pose thermalization problems in dry cryostats.
\subsection{Cryo probes and cabling}

In order to have high flexibility, QHE devices are typically installed into so called cryo probes or dip sticks. They are set up by a sample holder or carrier which is connected to a thin walled stainless steel tube with a connector box on top. The tube can be moved into the cryostat through a sliding seal. By this the loss of helium is minimized. To minimize the heat load to the cryogenic system the material used needs to have good mechanical properties and low thermal conductance as for example  stainless steel or fiberglass reinforced plastics. 

Using materials with low thermal conductivity for the needed coaxial cables contradicts the need of low electrical resistance. Since at least six coaxial lines are needed to connect one quantum Hall device, the heat load to the cryo system can be quite large when using standard coaxial cables. The introduced heat is already reduced when (sub) miniature coaxial cables are used. This is a possible solution e.g. for operation in a liquid helium Dewar in combination with a recovery system. Using such a coaxial cable with an characteristic impedance of \SI{75}{\ohm} will result in a cabling with low resistance and capacitance. Some care has to be taken to the material of the inner and outer conductor since pure copper can get brittle at cryogenic temperatures. This can be avoided using copper alloys or copper / steel compounds. 

For systems operated below liquid helium temperature (lambda cooler, \SI{1}{\kelvin} pot, $^3He$ systems, ...) or closed loop cryo cooler, the amount of heat introduced to the system needs to be further reduced. This is done by using e.g. brass, steel or stainless steel for the outer conductor and in some cases also for the inner conductor. This results in coaxial cables with  low thermal conductance but increases the electric resistance in the range of a few \SI{}{\ohm\per\m}.

\subsection{Sample holder}
\begin{figure}[tb]
\centering
\includegraphics[width=0.8\linewidth]{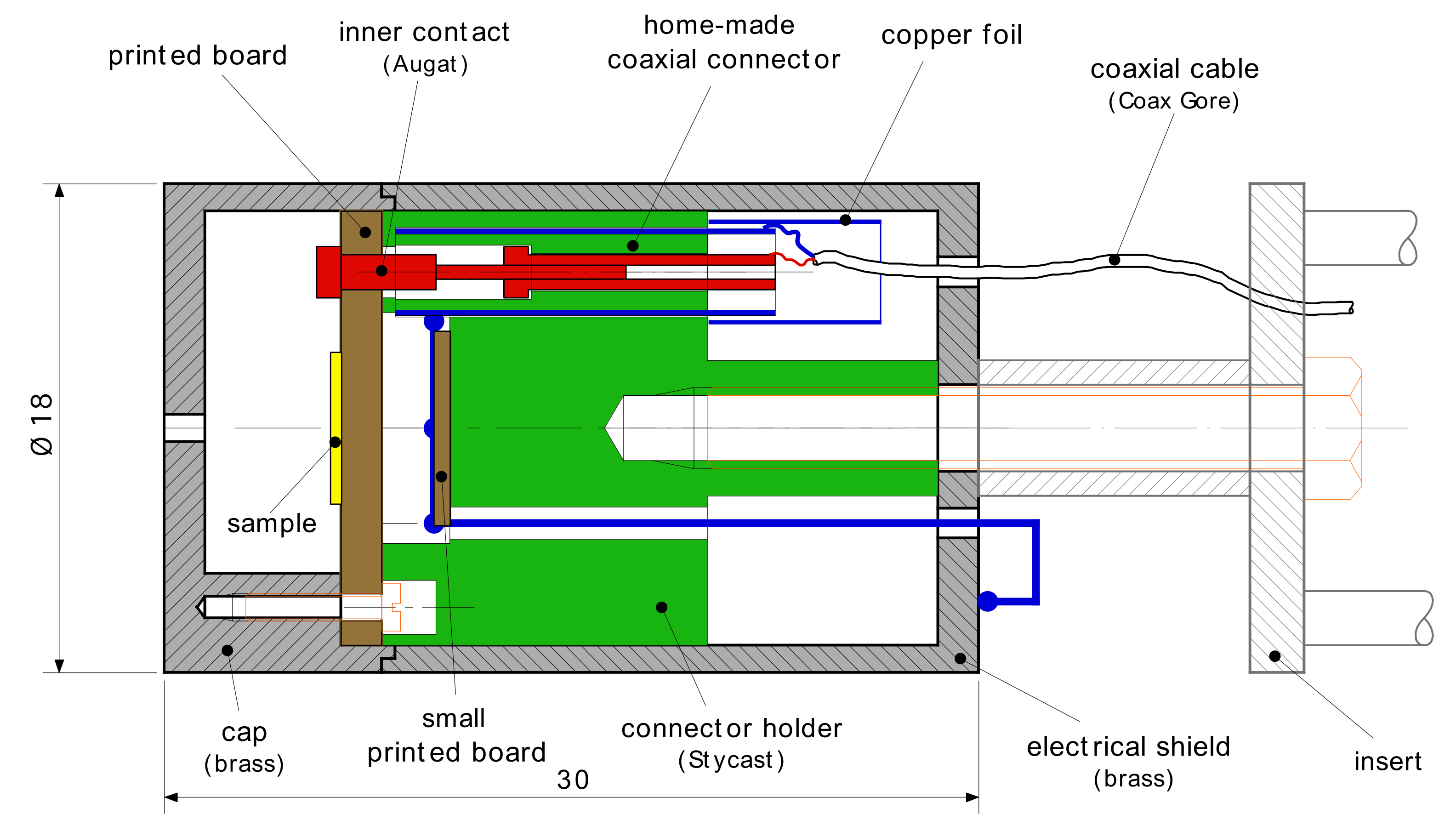}
\caption[EUROMET sample holder schematic]{Schematic diagram of the EUROMET sample holder.}
\label{fig:EUROMET_holder_scheme}
\end{figure}
\begin{figure}
\centering
\includegraphics[width=0.4\linewidth]{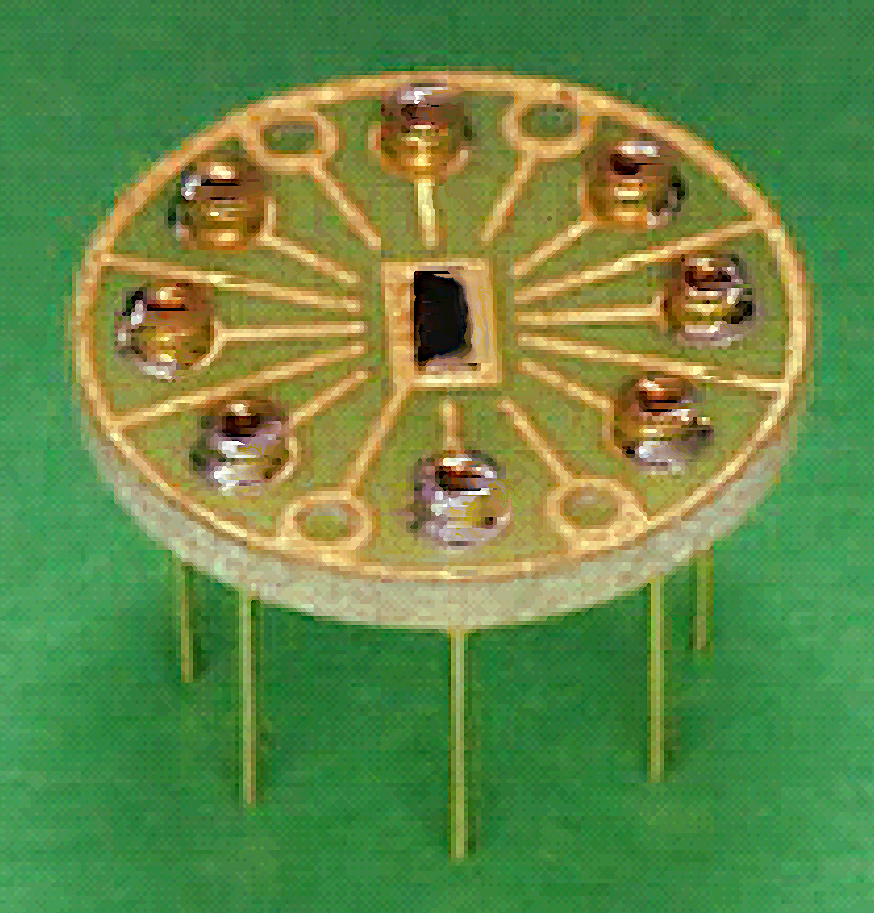}
\caption[EUROMET sample holder PCB]{The EUROMET holder printed circuit board where the QHE sample is bonded.}
\label{fig:EUROMET_holder_pic}
\end{figure}
\begin{figure}[tb]
\centering
   \begin{subfigure}[a]{0.5\textwidth}
         \centering
         \includegraphics[width=\textwidth]{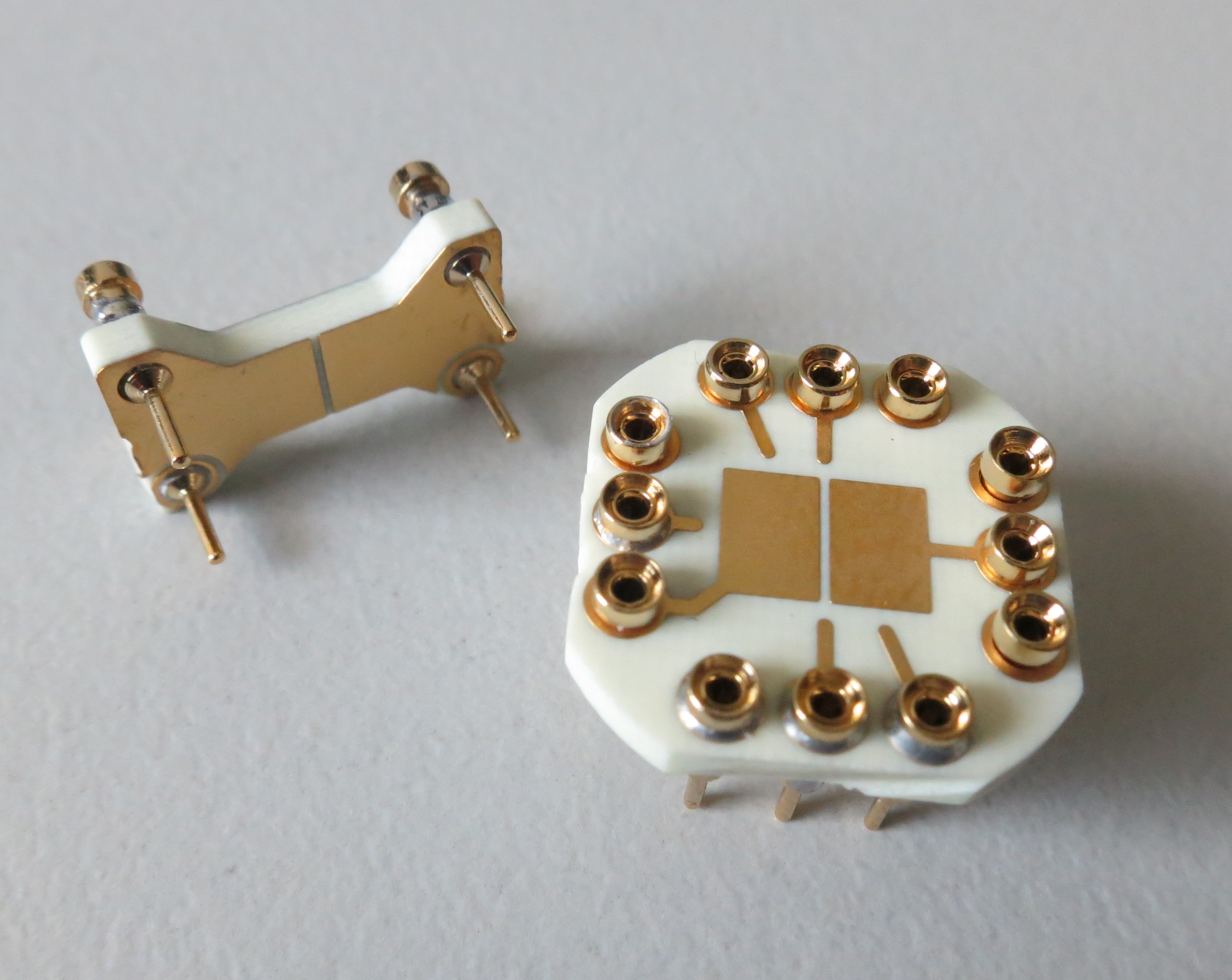}
         \caption{}
         \label{fig:to8_shielded}
     \end{subfigure}
   \begin{subfigure}[b]{0.8\textwidth}
         \centering
         \includegraphics[width=\textwidth]{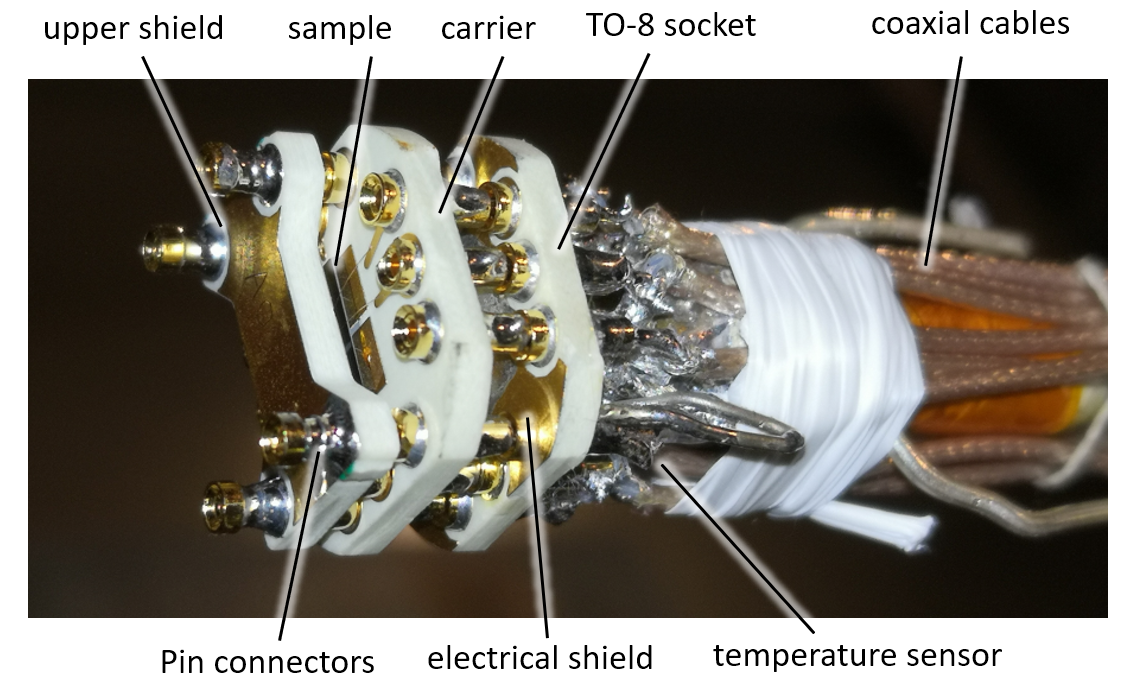}
         \caption{}
         \label{fig:to8_mounted}
     \end{subfigure}
\caption[TO-8 holder]{TO-8 holder, consisting of two PCBs with splitted shields. (\subref{fig:to8_shielded}) disassembled. (\subref{fig:to8_mounted}) with the QHR device mounted.}
\label{fig:to8}
\end{figure}
AC measurements ask for a definition of the quantum Hall device as a terminal-pair impedance standard. In order to achieve this, the normal QHR wiring exploited in dc (one conductor per contact) must be replaced by coaxial wiring. The sample holder must be designed to carry on the coaxial structure of the wiring as close as possible to the device. With the aforementioned connection point for all outer conductors.
\subsubsection{EUROMET holder}
The EUROMET holder was developed by the Swiss Federal Office of Metrology and Accreditation (METAS) in the framework of an EUROMET project~\cite{EUROMET540}. A schematic diagram of the holder is shown in Figure~\ref{fig:EUROMET_holder_scheme}, and a picture of the socket is shown in Figure~\ref{fig:EUROMET_holder_pic}.
\subsubsection{TO-8 shielded holder}
The TO-8 holder is adapted from the 12-pin version of the standardized TO-8 (\emph{transistor outline}) metal semiconductor package. It was developed by PTB and CMI~\cite{Kucera:2019} and implements the double-shielding technique of \cite{Kibble:2008}\footnote{A standard PCB material such FR4 was used. The introduction of modern materials, dedicated for high frequency applications and space technology, enable metrologists to design the holder with lower thermal expansion, conductivity, dissipation factor and lower dielectric constant, leading to lowering of several parasitic effects.}.
%
%
\clearpage
\section{Digital impedance bridges}
\begin{figure}[b]
\centering
\def\svgwidth{0.6\textwidth}
\begingroup%
  \makeatletter%
  \providecommand\color[2][]{%
    \errmessage{(Inkscape) Color is used for the text in Inkscape, but the package 'color.sty' is not loaded}%
    \renewcommand\color[2][]{}%
  }%
  \providecommand\transparent[1]{%
    \errmessage{(Inkscape) Transparency is used (non-zero) for the text in Inkscape, but the package 'transparent.sty' is not loaded}%
    \renewcommand\transparent[1]{}%
  }%
  \providecommand\rotatebox[2]{#2}%
  \newcommand*\fsize{\dimexpr\f@size pt\relax}%
  \newcommand*\lineheight[1]{\fontsize{\fsize}{#1\fsize}\selectfont}%
  \ifx\svgwidth\undefined%
    \setlength{\unitlength}{258.74780273bp}%
    \ifx\svgscale\undefined%
      \relax%
    \else%
      \setlength{\unitlength}{\unitlength * \real{\svgscale}}%
    \fi%
  \else%
    \setlength{\unitlength}{\svgwidth}%
  \fi%
  \global\let\svgwidth\undefined%
  \global\let\svgscale\undefined%
  \makeatother%
  \begin{picture}(1,0.62971801)%
    \lineheight{1}%
    \setlength\tabcolsep{0pt}%
    \put(0,0){\includegraphics[width=\unitlength,page=1]{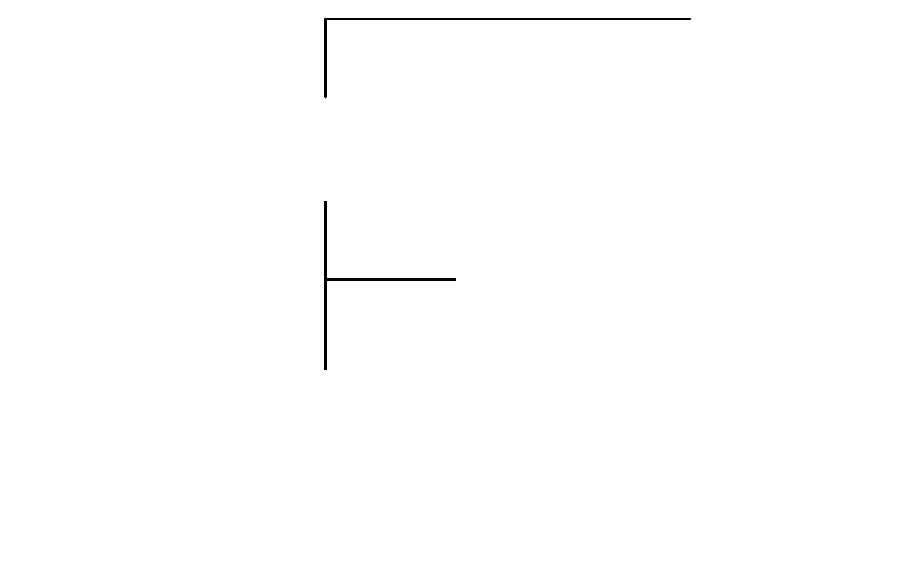}}%
    \put(0.27536466,0.44927918){\color[rgb]{0,0,0}\makebox(0,0)[rt]{\lineheight{0}\smash{\begin{tabular}[t]{r}$E_1$\end{tabular}}}}%
    \put(0,0){\includegraphics[width=\unitlength,page=2]{BridgePrinciple.pdf}}%
    \put(0.56522219,0.30435041){\color[rgb]{0,0,0}\makebox(0,0)[t]{\lineheight{0}\smash{\begin{tabular}[t]{c}D\end{tabular}}}}%
    \put(0.27536466,0.15942173){\color[rgb]{0,0,0}\makebox(0,0)[rt]{\lineheight{0}\smash{\begin{tabular}[t]{r}$E_2$\end{tabular}}}}%
    \put(0.8231954,0.44638069){\color[rgb]{0,0,0}\makebox(0,0)[lt]{\lineheight{0}\smash{\begin{tabular}[t]{l}$Z_1$\end{tabular}}}}%
    \put(0.82609397,0.15942173){\color[rgb]{0,0,0}\makebox(0,0)[lt]{\lineheight{0}\smash{\begin{tabular}[t]{l}$Z_2$\end{tabular}}}}%
    \put(0,0){\includegraphics[width=\unitlength,page=3]{BridgePrinciple.pdf}}%
  \end{picture}%
\endgroup%

\caption{Principle schematics of an impedance bridge.}
\label{fig:bridgeprinciple}
\end{figure}
\vspace{2mm}
The comparison of two impedance standards can be performed with an \emph{impedance bridge}, an instrument based on a measuring method invented by Christie~\cite{Christie:1833} and made popular by Wheatstone \cite{Wheatstone:1843}.
Figure~\ref{fig:bridgeprinciple} shows the principle schematic diagram of a generic impedance bridge. The two impedance standards $Z_1$ and $Z_2$ under comparison are connected in series. When a current $I$ flows through the impedances, two voltages $E_1 = Z_1 I$ and $E_2 = Z_2 I$ develop across $Z_1$ and $Z_2$. At equilibrium ($V_\mathrm{D} = 0, I_\mathrm{D} = 0)$ The impedance ratio is related to the voltage ratio by the equation
\begin{equation}
\frac{Z_1}{Z_2} = -\frac{E_1}{E_2}.
\end{equation}

In the Wheatstone bridge voltages $E_1$ and $E_2$ are provided by a single source, using a resistive divider.

In transformer bridges \cite{Kibble:1984,Awan:2010} the voltages $E_1$ and $E_2$ are generated by an inductive voltage divider; the nominal $E_1/E_2$ ratio is determined (and very close numerically) to the turns ratio, and can be calibrated to a very high accuracy. Transformer bridges allow extremely accurate ratio measurements, reaching uncertainties of parts in \num{E9} \cite{Schurr:2009}, but are large and complex electrical networks, can measure only impedance ratios very close to the limited set of nominal ratios fixed by construction, and their frequency bandwidth is limited.

\subsection{Digital bridges} 
Digital bridges are impedance bridges that make extensive use of mixed-signal electronic devices, either analog-to-digital converters (ADC) or digital-to-analog converters (DAC) and have digital representations of the voltage and current waveforms in the bridge mesh.

The concept of digital bridges dates back to decades ago \cite{Helbach:1983, Helbach:1987, Tarach:1993}, but only more recently the performances of ADCs and DACs improved to underpin the implementation of primary impedance bridges. With respect to transformer impedance bridges, digital bridges have simpler electrical networks, are less expensive and easier to automate.
\subsection{Digitally-assisted bridges}
Digitally-assisted bridges \cite{Cabiati:1985, Tarach:1993, Waltrip:1995, Muciek:1997} are digital bridges that employ electromagnetic components, transformers and inductive voltage dividers, as ratio standards. The bridge is energized by a large-amplitude signal; a number of auxiliary signals (voltages and currents) of smaller amplitude are employed to achieve the main equilibrium (that gives the bridge reading) and the auxiliary equilibria necessary to set the impedance standards in the proper defining conditions. The main and auxiliary signals are generated by digital synthesis with DACs. 

The accuracy of digitally-assisted bridges is guaranteed by the electromagnetic ratio standard, hence the bridge performances - and the corresponding limitations (fixed set of ratios and frequencies available) are similar to those of the traditional transformer bridges. The digital source employed must guarantee a limited harmonic content and noise, but the requirements on the accuracy and stability are moderate (in the \num{E-4} range). 

An implementation with digitally-assisted bridges of capacitance realisation from the dc quantum Hall effect is presently operating \cite{Callegaro:2010}. 
\subsection{Electronic fully-digital bridges}
\label{sec:elecfdbridges}
\begin{figure}[t]
	\center
	\includegraphics[width=0.5\columnwidth]{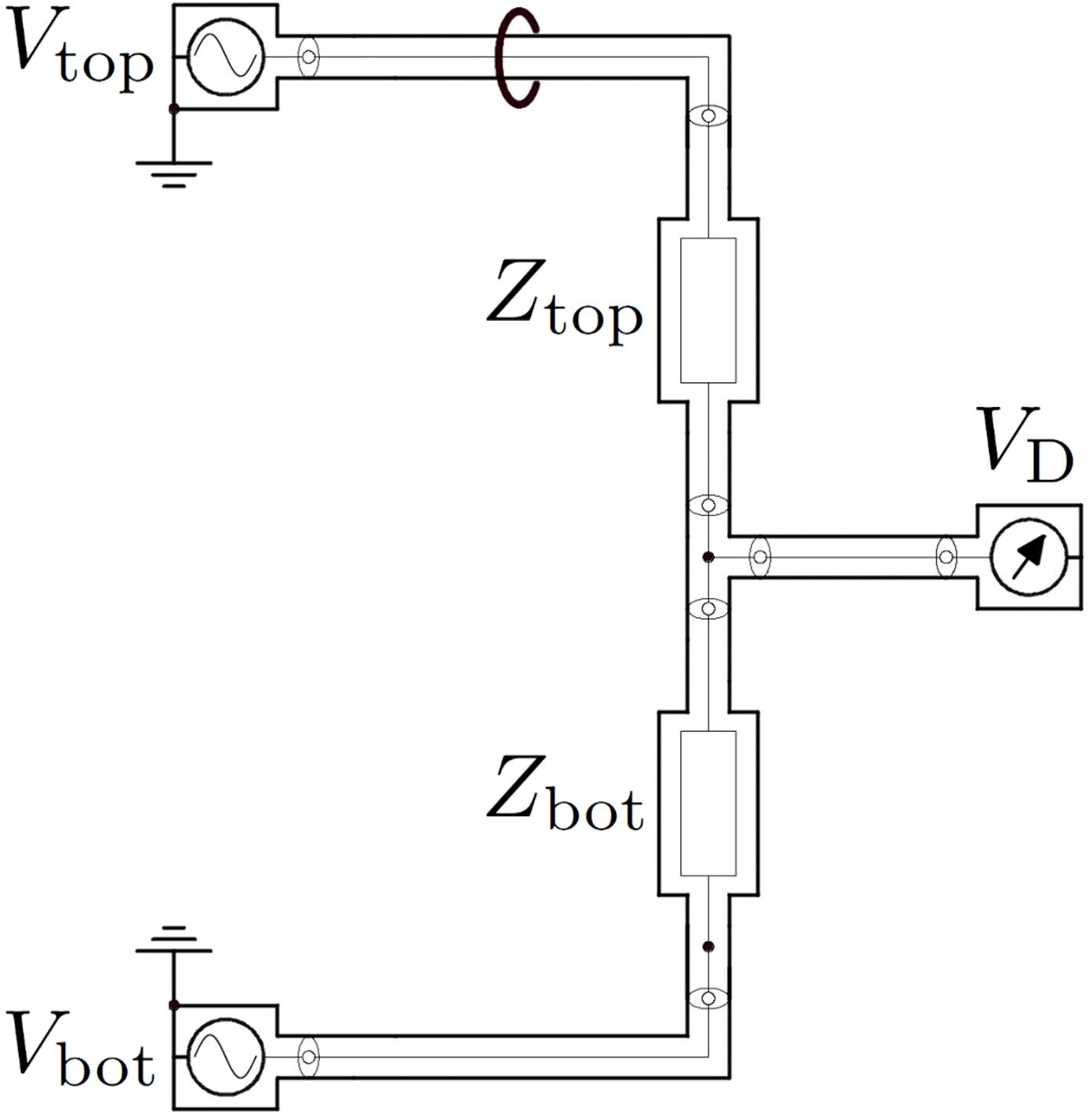} 
	\caption[Two terminal-pair electronic source bridge]{Simple schematic of a two terminal-pair electronic source bridge. The amplitudes and phases of the sources $V_{\rm top}$ and $V_{\rm bot}$ are adjusted to obtain $V_{\rm D}=0$ (the main balance), then $Z_{\rm bot}/Z_{\rm top}=-V_{\rm bot}/V_{\rm top}$.}
	\label{fig:SolidStateBridge}
\end{figure}
\begin{figure}[t]
	\center
    \includegraphics[width=0.9\columnwidth]{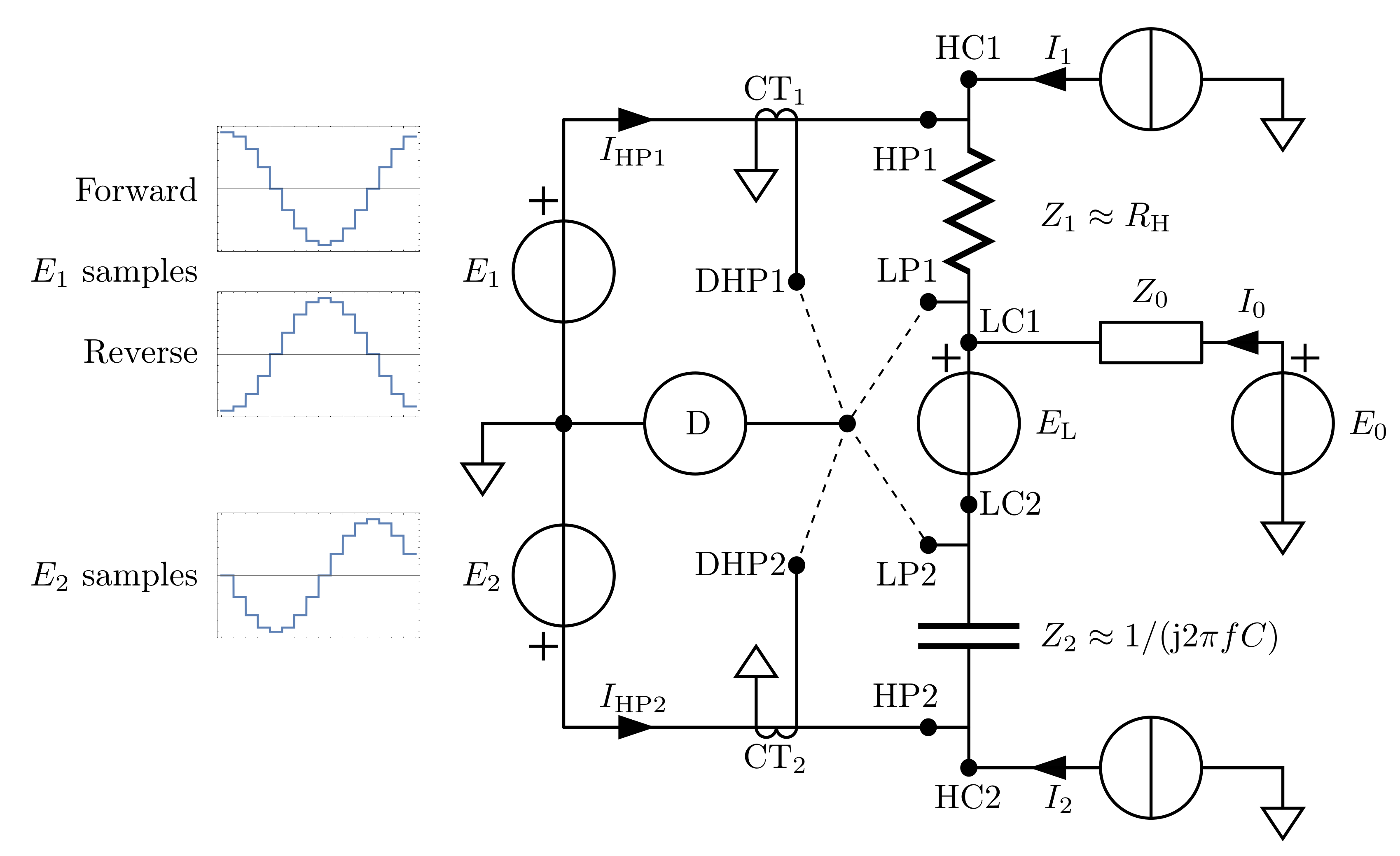} 
	\caption[Fully-digital electronic bridge for $R-C$ comparisons]{Simplified principle schematic of a fully-digital electronic bridge suitable for a direct $R-C$ comparison between standards defined as four terminal-pair impedances. $Z_1 = R_\mathrm{H}$ and $Z_2 = 1/(j \omega C)$ are the impedances under comparison; $\omega$ is the bridge operating frequency, chosen so that $\omega R C = 1$; $E_1$ and $E_2$ are the main bridge voltages; $I_1$ and $I_2$ are the current sources balancing $I_\mathrm{HP1}$ and $I_\mathrm{HP2}$; $E_\mathrm{L}$ is the voltage source balancing the difference $V_\mathrm{LP1}-V_\mathrm{LP2}$; CT1 and CT2 are current transformers measuring the currents $I_\mathrm{HP1}$ and $I_\mathrm{HP2}$, respectively; the voltage source $E_0$ and the impedance $Z_0$ constitute an auxiliary injection arm to fine-tune the bridge balance; and D is a synchronous detector that can be connected, in turn, to the detection terminals LP1, LP2, DHP1 and DHP2. The diagrams on the left represent example waveform samples: the samples of $E_1$ are changed in sign between the forward and reverse configurations; the samples of $E_2$ are instead kept fixed. After Ref.~\cite{Marzano:2020}, courtesy of the authors.}
	\label{fig:INRIMbridge}
\end{figure}
Electronic fully-digital bridges \cite{Bachmair:1980, Helbach:1983} are voltage ratio bidges where the bridge voltage ratio standard is based on the linearity properties of DACs (or ADCs). The digital source employed must therefore comply with strict requirements of accuracy and stability. Since sampling and digital processing allow to generate (or measure) arbitrary voltage ratios, frequencies and phase differences, and frequencies, fully-digital bridges overcome the intrinsic limitations of transformer bridges, either traditional or digitally-assisted. 

A simple implementation of a two terminal-pair digital bridge \cite{Callegaro:2015} is shown in Figure~\ref{fig:SolidStateBridge}. Two digital voltage sources supply the measuring current to the two impedances standards, which are connected in series. The amplitude and phase of one of the sources is adjusted \cite{Dutta:2001} until the voltage measured at the node between the two impedances is set to zero ($V_D=0$). The impedance ratio determination is completely relying on the the agreement between the settings of the generators and the actual voltages applied to the impedances. 

It is possible to perform two different measurements by reversing the top and bottom standards. Averaging the readings obtained in the two configurations  allows for a significant reduction of the error related to a possible asymmetry between the two channels of the  source~\cite{Callegaro:2015,Kucera:2018}. The remaining uncertainty components are related to the non-linearities\cite{Helbach:1983}, gain and phase stability~\cite{Muciek:1997,Kampik:2016} and loading effect of the parasitic stray admittances~\cite{Callegaro:2015}. 

Some published implementations of digital bridges still rely on inductive voltage dividers for the voltage ratio adjustment \cite{Muciek:1997,Sedlacek:2005a,Sedlacek:2009}, but they can be anyway used to perform comparisons of any kind of impedances, like $R$-$C$ comparisons \cite{Bachmair:1980,Ramm:1985a} or $L$-$C$ comparisons \cite{Cabiati:1985}. Four terminal-pair versions of digital bridges have also been successfully developed \cite{Sedlacek:2005a,Sedlacek:2009,Lan:2012, Kucera:2018}.

The operating frequency range of these bridges starts from a few tens of \si{\hertz} \cite{Ramm:1985a} or even lower \cite{Lan:2012} to \SI{10}{\kilo\hertz} \cite{Sedlacek:2009}. The uncertainty on the generated voltage ratio ranges from a few parts in \num{E5}~\cite{Muciek:1997} down to less than 1 part in \num{E6}~\cite{Bachmair:1980, Trinchera:2009}.

Specialized fully-digital bridges suitable for a direct realisation of the farad in terms of the quantized Hall resistance can be conceived. On example of such a bridge is given in ~\cite{Marzano:2020}; Figure~\ref{fig:INRIMbridge} shows its working principle. The restriction of the operational parameters to the condition $\omega R C = 1$ minimizes the contribution of DAC nonlinearty to the bridge uncertainty.

\subsection{Josephson bridges}
\label{sec:josephsonbridges}
The accuracy limitation of electronic DACs employed in a fully-digital bridge can be overcome by employing Josephson DACs. These are composed of integrated circuits including thousands of Josephson junctions in series, called Josephson arrays, driven by room-temperature electronics. 

The two main types of Josephson DACs available give rise to two different waveform synthesis methods:
\begin{description}
\item[Programmable Josephson Voltage Generators (PJVS)] work like binary-weighed DACs: the array is divided into segments composed of a binary sequence (1, 2, 4, 8, \ldots) of junctions in series. The segment can generate, when driven by a proper dc bias current, a positive, zero or negative quantized voltage. The voltage is proportional to the number of junctions in each segment. 

In static bias conditions of bias the output of a binary array is a quantized voltage 
\begin{equation*}
V = b \frac{f}{K_\mathrm{J}}, 
\end{equation*}

where $b$ is an integer dependent on the code selected to drive the binary segments, $f$ is the frequency of the biasing microwave, and $K_\mathrm{J}$ is the Josephson constant. 

The limitations of binary Josepshon array DACs are the resolution in bits, typically $10$ to $16$,  and thus which in turn is dependent by the number of integrated Josephson junctions, and the settling time. The latter limits the maximum operating frequency to achieve quantum performance to the \si{\kilo\hertz} at most. Josephson voltage ratio bridges with relative accuracies in the \num{E-8} range have been published \cite{Lee:2011}.

\item[Josephson array waveform synthesizers (JAWS)] are based on arrays of $N$ junctions in series, driven by a sequence of microwave pulses \cite{Benz:1996}. The array transform each driving pulse in a corresponding voltage impulse of quantized amplitude $V \mathrm{d}t = N K_\mathrm{J}^{-1}$. The  rf pulses are rejected by LP filters, and a voltage $V = p N K_\mathrm{J}^{-1}$ appears at the output, where $p$ is the pulse rate. When operating the array with positive and negative pulses and a pulse sequence where the pulse rate varies in time, arbitrary wave forms can be generated. 

Pulse-driven array DACs are typically drive by pulse pattern generators with generation rates in the \si{\giga\hertz} range, and allow to generate high-resolution sinewaves with frequencies up the \si{\mega\hertz} range. The major limitation is related to the maximum output voltage, although recently multi-chip generators with $> \SI{1}{\volt}$ output range \cite{Kieler:2021} have been achieved. 
\end{description}
%


\subsubsection{Programmable Josephson Bridges}
\begin{figure}
	\center
	\includegraphics[width=0.7\columnwidth]{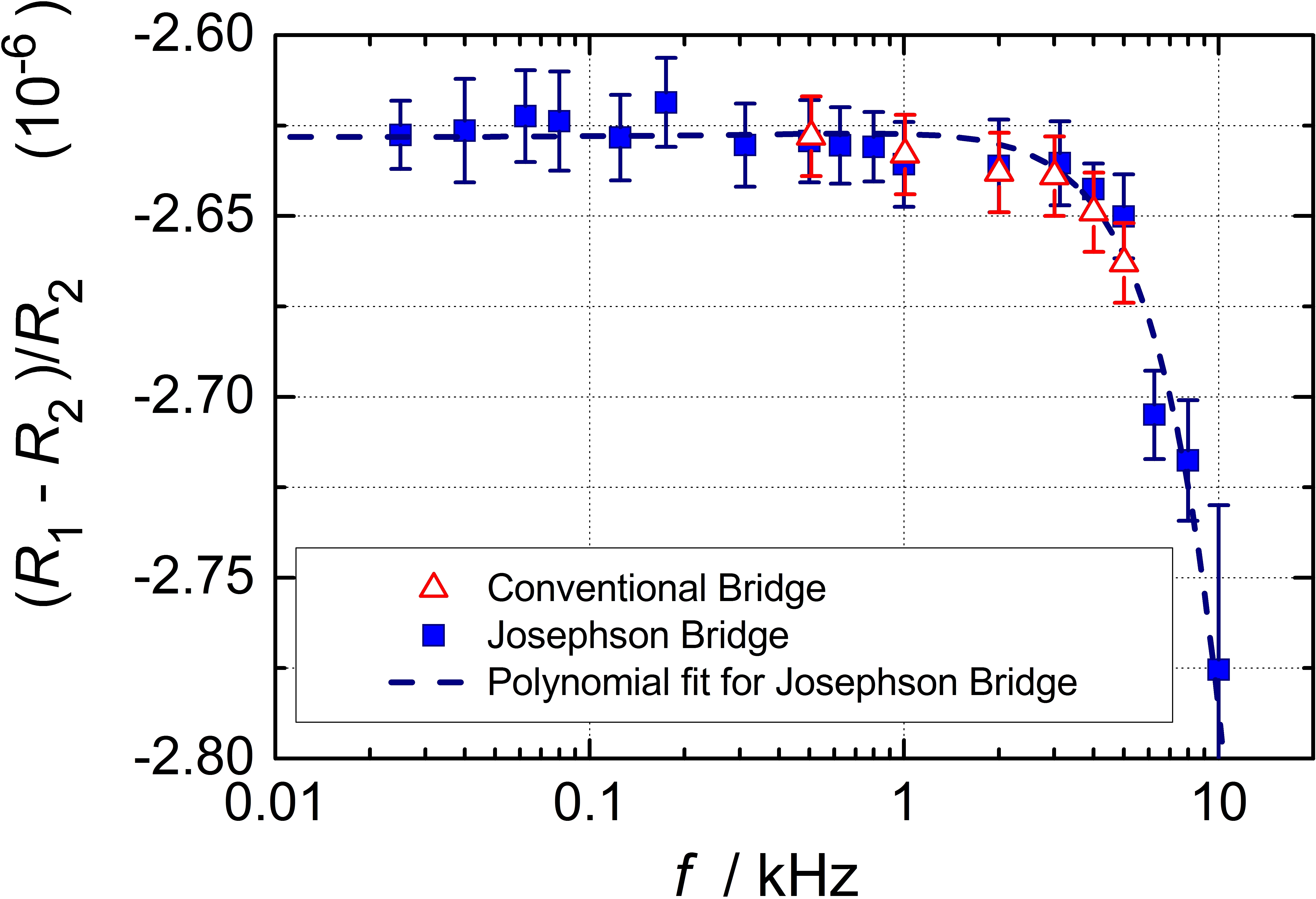}
	\caption[Relative difference between two \SI{10}{\kilo\ohm} resistors versus frequency]{Measurement of the relative difference between two \SI{10}{\kilo\ohm} resistors versus frequency, performed using traditional transformer based bridge and the 2TP PJVS based bridge (after \cite{Lee:2010a}, courtesy J. Lee).}
	\label{PJVS_Bridge_2TP_Results}
\end{figure}
Programmable Josephson Voltage Standards (PJVS) generate stepwise approximated ac waveforms, that can be used to generate accurate voltage ratios. The first bridge based on PJVS synthesized voltages, a two terminal-pair bridge, was demonstrated in 2010 \cite{Lee:2010a}. The bridge was employed to compare two 10~k$\Omega$ resistances, reaching an  accuracy comparable to that of traditional transformer-based bridges. Figure \ref{PJVS_Bridge_2TP_Results} shows that the agreement between the measurements performed with a traditional 2TP transformer bridge and the PJVS bridge is around a few parts in \num{E8}, well within the measurement uncertainties. One can appreciate that thePJVS bridge measurements were performed over a  wider frequency range (from \SI{25}{\hertz} to \SI{10}{\kilo\hertz}). This 2TP bridge was also used to compare two \SI{100}{\pico\farad} capacitance standards \cite{Palafox:2012} with a relative accuracy of \num{2E-8} at \SI{1}{\kilo\hertz}. 

The bridge was then employed, with a ratio of 10:1, to compare two capacitors (\SI{10}{\pico\farad} to \SI{100}{\pico\farad}) in the frequency range \SI{25}{\hertz} and \SI{20}{\kilo\hertz} \cite{Palafox:2014}. The uncertainty  was \num{3E-7}, with full agreement with traditional methods for frequencies around \SI{1}{\kilo\hertz}. The bridge was equipped with \SI{10}{\volt} arrays \cite{Palafox:2014}, which reduced the signal to noise ratio. Unfortunately, at \SI{10}{\volt}, the helium consumption increased to a level which is quoted as "prohibitive" in \cite{Palafox:2014}.    

The 2TP programmable Josephson bridge was then upgraded to a 4TP version, in order to compare low-valued impedances. The first set of measurements performed was the comparison of two \SI{10}{\kilo\ohm} resistance standards over the frequency range \SI{20}{\hertz} to \SI{10}{\kilo\hertz} \cite{Lee:2011}. The outcome of the measurement show a systematic offset of \num{6E-8}, which is larger than the (type A) measurement uncertainty of a few parts in \num{E9}. Moreover, the frequency dependence shows the wrong curvature above \SI{3}{\kilo\hertz}. Such behavior was tracked back to the large harmonic content of the PJVS waveform, which is difficult to cope with in this type of bridge configuration. The 4TP bridge was further refined  recently \cite{Hagen:2017}, but the improvement was not considered satisfactory. Quoting \cite{Hagen:2017}: "These results clearly show that further ideas are needed to setup a 4TP Josephson impedance bridge based on PJVS". 

The 2TP bridge performance was also tested in the measurement of two unlike impedances, a \SI{12.9}{\kilo\ohm} resistor versus a \SI{10}{\nano\farad} capacitor at the frequency of \SI{1233}{\hertz} \cite{Palafox:2012}. The type A uncertainty was \num{1.5E-6}; a comparison with a traditional measurement agrees within such uncertainty. Again, the limitation of the system was due to the large harmonic content of the PJVS waveforms.

In summary, the Josephson impedance bridge based on a PJVS has certainly played an relevant role in impedance metrology research, by showing that Josephson arrays can be used in impedance bridges to achieve competitive uncertainties. However, the large harmonic content of the PJVS waveforms, intrinsic to the working principle of the PJVS, considerably limits the application area of this type of bridges. In this sense the Dual Josephson Impedance Bridge, based on JAWS, represents a much more promising approach.  

\subsubsection{Dual Josephson Impedance Bridge (DJIB)}
\begin{figure}[p]
	\center
	\includegraphics[width=0.95\columnwidth]{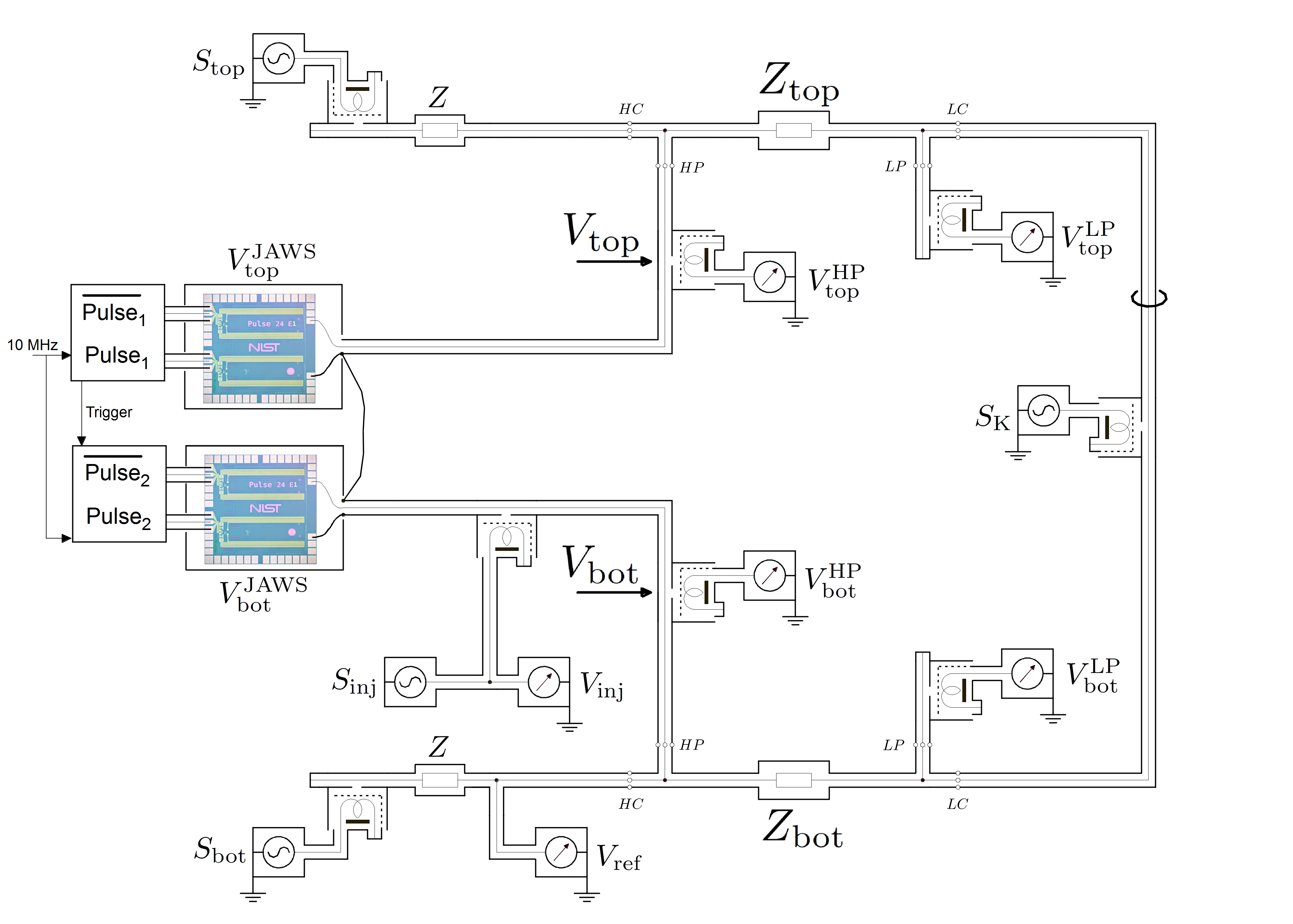}
	\caption[Simplified DJIB bridge schematic]{Simplified coaxial schematic of a four-terminal pair DJIB bridge circuit. The bridge by setting $V_{\rm top}^{\rm HP}$, $V_{\rm bot}^{\rm HP}$, $V_{\rm top}^{\rm LP}$ and $V_{\rm bot}^{\rm LP}$ to zero; this is achieved by adjusting the amplitude and the phase of the bottom JAWS source, as well as the voltages $S_{\rm top}$, $S_{\rm bot}$, and $S_{\rm K}$, which give the four-terminal-pair definition of the two impedance standards. In this condition, the impedance ratio $Z_{\rm bot}/Z_{\rm top}$ is equal to the voltage ratio $-V_{\rm bot}/V_{\rm top}$. The difference between the DJIB and the digital assisted bridge \cite{Overney:2016} is that the accurate and stable voltage ratio is generated using two JAWS systems instead of a ratio transformer. The full description of the bridge can be found in \cite{Overney:2016b}.}
	\label{fig:DJIBschematic}
\end{figure}
Josephson arbitrary waveform synthesizers (JAWS) are digital-to-analog converters that generate quantum-accurate distortion-free voltage wave forms over frequencies in the \si{\hertz} to \si{\mega\hertz} frequency range. By combining and synchronizing two such JAWS systems enables generation of quantum-accurate, calculable voltages with arbitrary ratios and arbitrary relative phase angles.

The two voltage sources required by a DJIB are provided by two independent pulse-driven JAWS systems operated in either a single or two separate dewars of liquid helium. The setup used at METAS each JAWS chip (NIST) include four arrays of 12800 double-stacked Josephson junctions (JJs) each, connected in series by on-chip superconducting traces to produce a voltage of 1~V. The PTB system, in a typical configurations, employs two independent JJ arrays having up to 12000 JJ. The clock signals of the JAWS system and of the other components of the DJIB are all locked to a 10~MHz reference frequency signal. The phase matching of the two JAWS is ensured because the two pulse generators share a single \SI{14.4}{\giga\hertz} clock. More details for both systems are given in \cite{Flowers-Jacobs:2016, Bauer:2021}. The performances of the DJIB relies on the stability, linearity, and tunability of the two JAWS systems.  

Figure \ref{fig:DJIBschematic} shows a simplified schematic of the DJIB, which was developed at METAS to perform high accuracy comparisons of the four terminal-pair impedances $Z_{\rm top}$ and $Z_{\rm bot}$. The working principle of the DJIB is similar to that of the digitally assisted bridge (DAB) described in \cite{Overney:2016} and is given in \cite{Overney:2016b}. The principal difference between the two bridges is that in the DJIB the accurate and stable voltage ratio is generated using two JAWS sources, whereas in the DAB it is achieved with a ratio transformer. Therefore, in the DJIB the amplitudes and the phases of $V_{\rm top}^{\rm JAWS}$ and $V_{\rm bot}^{\rm JAWS}$ can be independently set to any desired value, hence making the comparison of arbitrary impedances possible.

A first test performed with the DJIB consisted in the measurement of the relative frequency dependence of two resistance standards, $Z_{\rm 12k9}^{\rm B}$ and $Z_{\rm 12k9}^{\rm A}$ (taken as a reference). The lower part of figure~\ref{fig_resistanceRatio} displays the outcome of the measurements, performed between \SI{1}{\kilo\hertz} and \SI{20}{\kilo\hertz} at the voltage \SI{1}{\volt} rms. The points give the values measured using the DJIB, the solid line is a quadratic fit to the measurements performed with the DAB.

Figure \ref{fig_resistanceRatio}, upper part, gives the difference between the values measured with the DJIB and those measured with the DAB are shown. The gray zone represents the combined (1-$\sigma$) uncertainty of the DAB \cite{Overney:2016}, while the bars correspond solely to the Type A uncertainties of the DJIB measurements. The DJIB measurements were repeated a number of times over a few days;  corrections for the small drift in the dc resistance were applied. At these particular frequencies, the residual spread of the results is slightly  larger than the Type A uncertainty. Such deviations indicates that some systematic effects remain to be investigated. Nevertheless, there is an agreement between the results obtained with the DJIB and the DAB better than 0.1~$\mu \Omega/\Omega$, confirming the potential  of the JAWS sources when implemented in an impedance bridge.

In addition, consistency check to compare different stable unknown impedance standards were performed.  Figure \ref{fig:ConsistencyCheck} shows the residuals $|\Delta|$, smaller than \SI{0.5}{\micro\ohm\per\ohm}, which is ten times better than the results obtained using the sampling bridge.
\begin{figure}[t]
	\center
	\includegraphics[width=0.7\columnwidth]{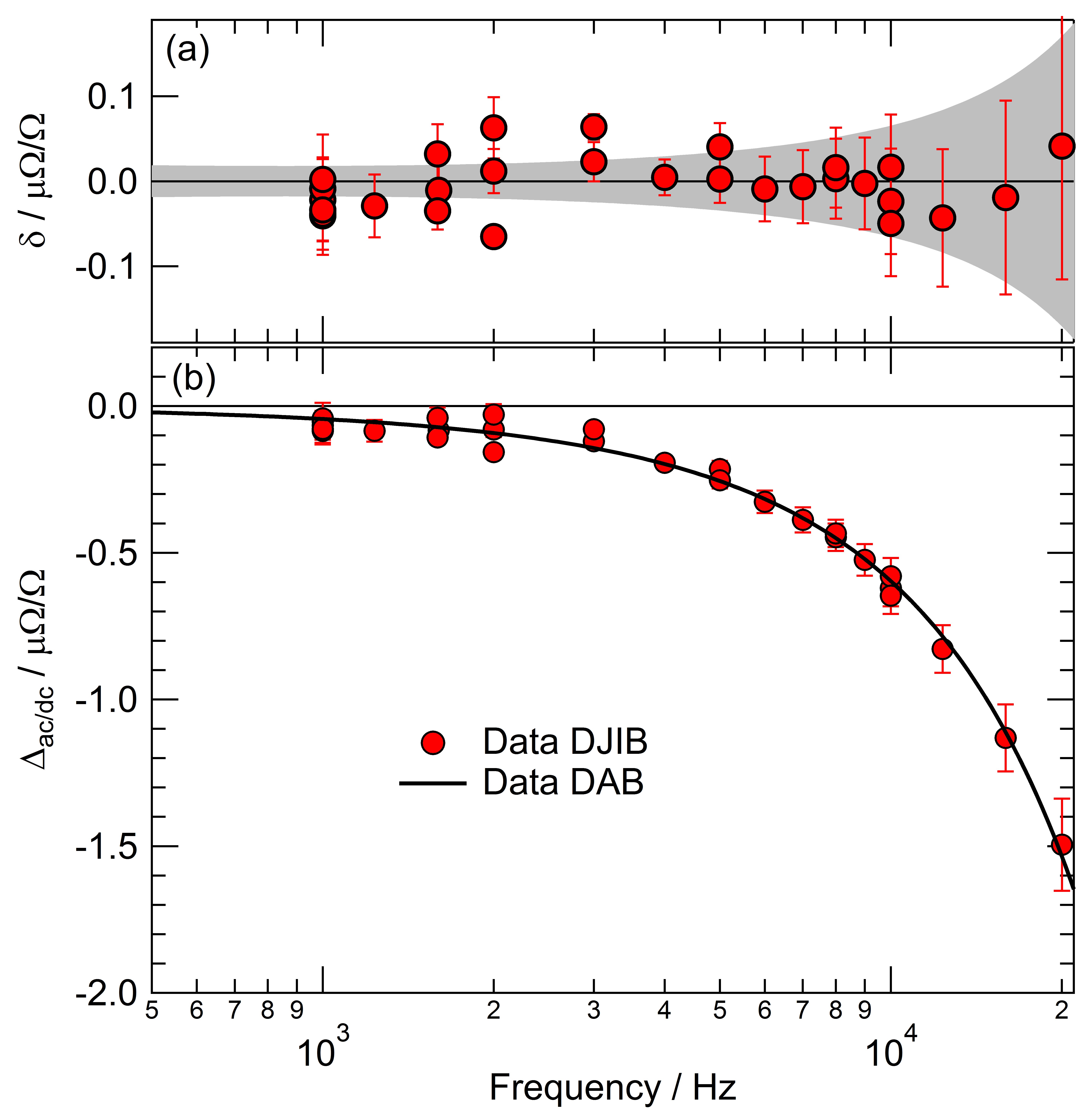}
	\caption[DJIB versus DAB bridge measurements]{The bottom plot (b) shows the frequency dependence, $\Delta_{ac/dc}$, of the resistance $Z_{\rm 12k9}^{\rm B}$ measured with the DJIB (symbols) and with the DAB (solid line). The top plot (a) shows the difference $\delta$ between the DJIB and DAB results. The errors bars correspond to the Type A uncertainty of the DJIB measurement. The gray zone represents the combined (k=1) uncertainties for the measurements made with the DAB \cite{Overney:2016}.}
	\label{fig_resistanceRatio}
\end{figure}
\begin{figure}[t]
	\center
	\includegraphics[width=0.5\columnwidth]{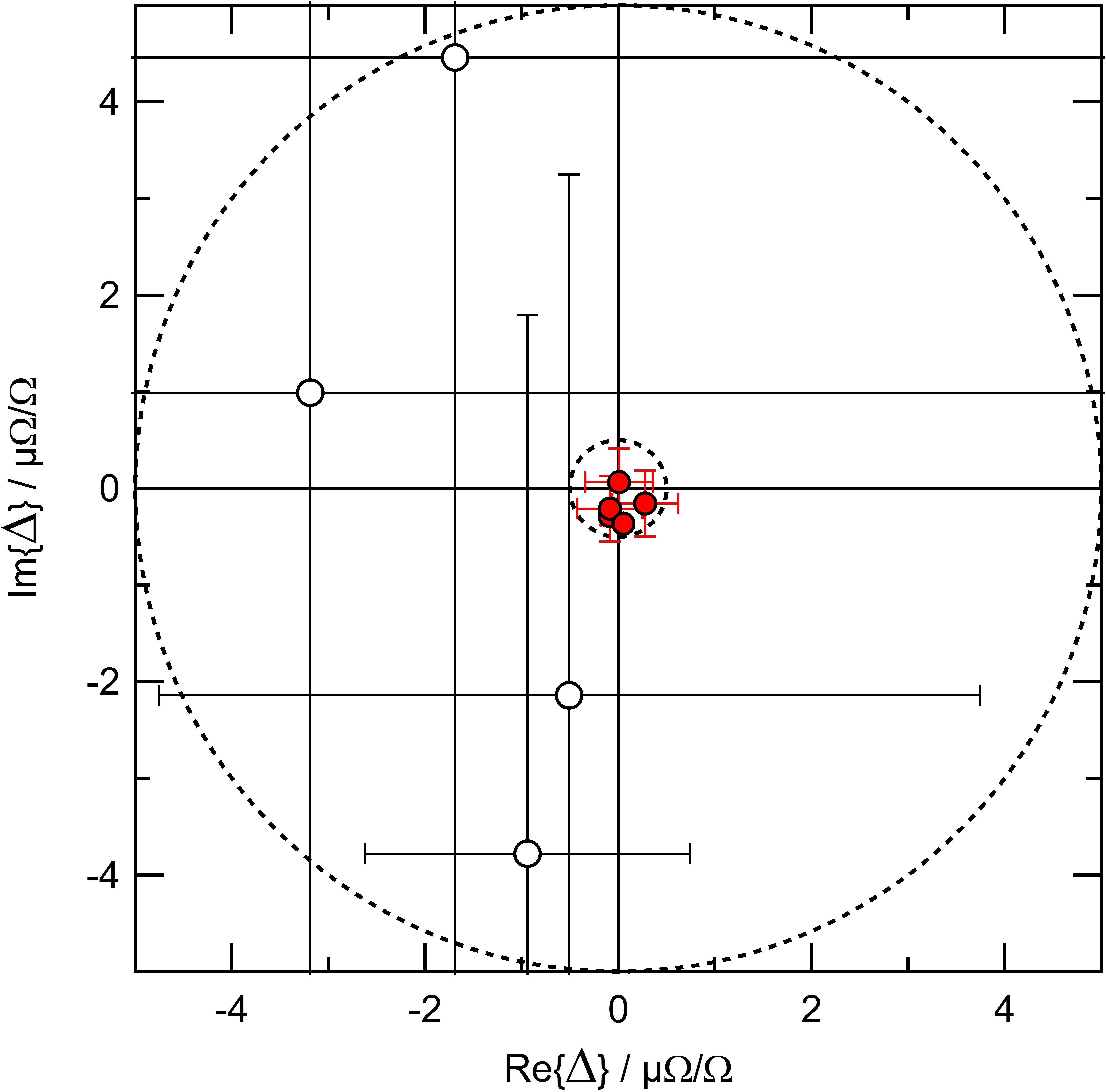}
	\caption[DJIB consistency check]{Residuals, $|\Delta|$, of the consistency checks for the sampling bridge \cite{Overney:2011c} (open circles) and the DJIB \cite{Overney:2016b} (solid circles). Dashed line circles correspond to $|\Delta|<5\ \mu\Omega/\Omega$ and \mbox{$|\Delta|<0.5\ \mu\Omega/\Omega$}.}
	\label{fig:ConsistencyCheck}
\end{figure}

Another four terminal-pair Josephson impedance brige, developed at PTB, is shown in Figure~\ref{PTB_JAWS_4TP_Bridge_Schema} and described in full details in \cite{Bauer:2021}.  The two impedances to be compared are biased by the voltages $U_{1}$ and $U_{2}$ provided by two JJA. Both arrays are driven by a pulse pattern generator with has two independent but synchronized memories sharing the same rf clock. The phase angle between $U_{1}$ and $U_{2}$ can be varied by changing the delay between both memories with a maximum resolution of \SI{250}{fs}. To balance the bridge, one of the voltage amplitudes and the phase angle between both voltages are adjusted until detector D is minimized. For a quadrature bridge with a resistance $R$ and a capacitance $C$, the real part of the bridge equation becomes $\omega RC = U_{1}/U_{2}$ with a 90~degree phase angle between $U_{1}$ and $U_{2}$. 

In Figure \ref{PTB_JAWS_4TP_Bridge_Schema} , the DJIB is set up with a QHR and a \SI{10}{\nano\farad} capacitance standard, resulting in a nominal signal frequency of \SI{1233}{\hertz}\. To eliminate the influence of the contact and the lead resistance, the QHR is connected to the bridge using a triple series connection \cite{Delahaye:1993b}. 

The quantum based voltage generated by each JJA is affected by the output cable impedances \cite{Filipski:2011,Brom:2012} which are matched to mitigate their influence. This influence is further reduced by interchanging the position of the measured impedances inside the bridge setup at the level of the current detection $D_1$ and $D_2$ (see Figure \ref{PTB_JAWS_4TP_Bridge_Schema}).

\begin{figure}[t]
	\center
	\includegraphics[width=0.7\columnwidth]{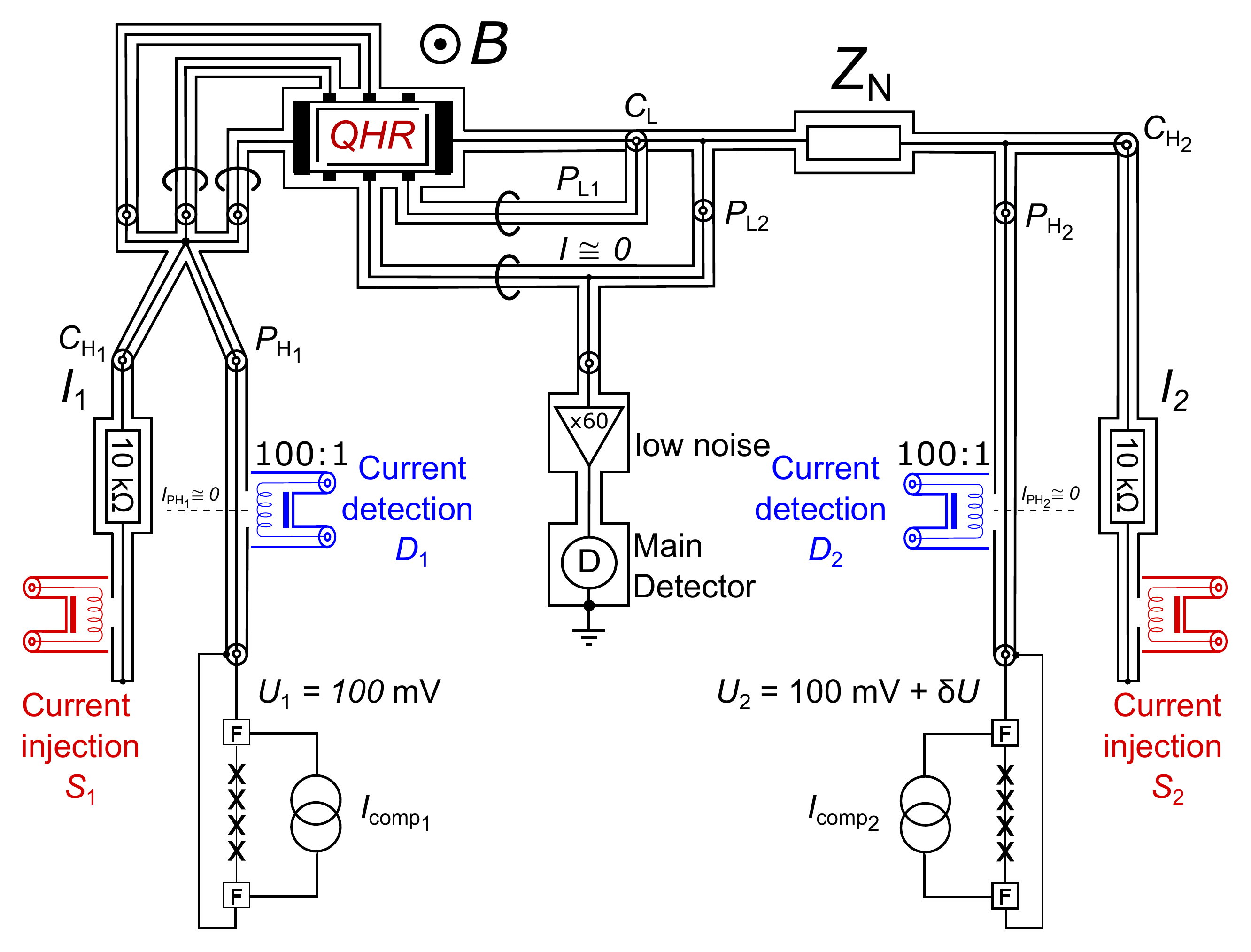}
	\caption[\SI{10}{\nano\farad} versus QHR measurement]{Schematic overview of the measurement setup for the quadrature measurements with a \SI{10}{\nano\farad} capacitance standard and a QHR (after \cite{Bauer:2021}, courtesy S. Bauer).}
	\label{PTB_JAWS_4TP_Bridge_Schema}
\end{figure}

With this universal impedance bridge a large variety of like and unlike impedacnes ca be compared with low uncertaintes and a measurement noise which is as lo as expected by theory. 
\clearpage
\section{Maintaining a capacitance scale with the AC QHE in graphene}
\subsection{Calibration and traceability}
The concepts of calibration and traceability are defined in the \emph{International Vocabulary of Metrology }~\cite{VIM}, here quoted:
\begin{description}
\item[2.39 Calibration] operation that, under specified conditions, in a first step, establishes a relation between the quantity
values with measurement uncertainties provided
by measurement standards and corresponding
indications with associated measurement uncertainties
and, in a second step, uses this information
to establish a relation for obtaining a measurement
result from an indication.
\item[2.41 Metrological traceability] property of a measurement result whereby the result can be related to a reference through a
documented unbroken chain of calibrations, each contributing to the measurement uncertainty.

(NOTE 2) Metrological traceability requires an established
calibration hierarchy.

(NOTE 3) Specification of the reference must include the
time at which this reference was used in establishing the
calibration hierarchy, along with any other relevant
metrological information about the reference, such as
when the first calibration in the calibration hierarchy was
performed.
\end{description}
\subsection{Traceability chains}
\begin{figure}[t]
	\center
	\includegraphics[width=0.6\columnwidth]{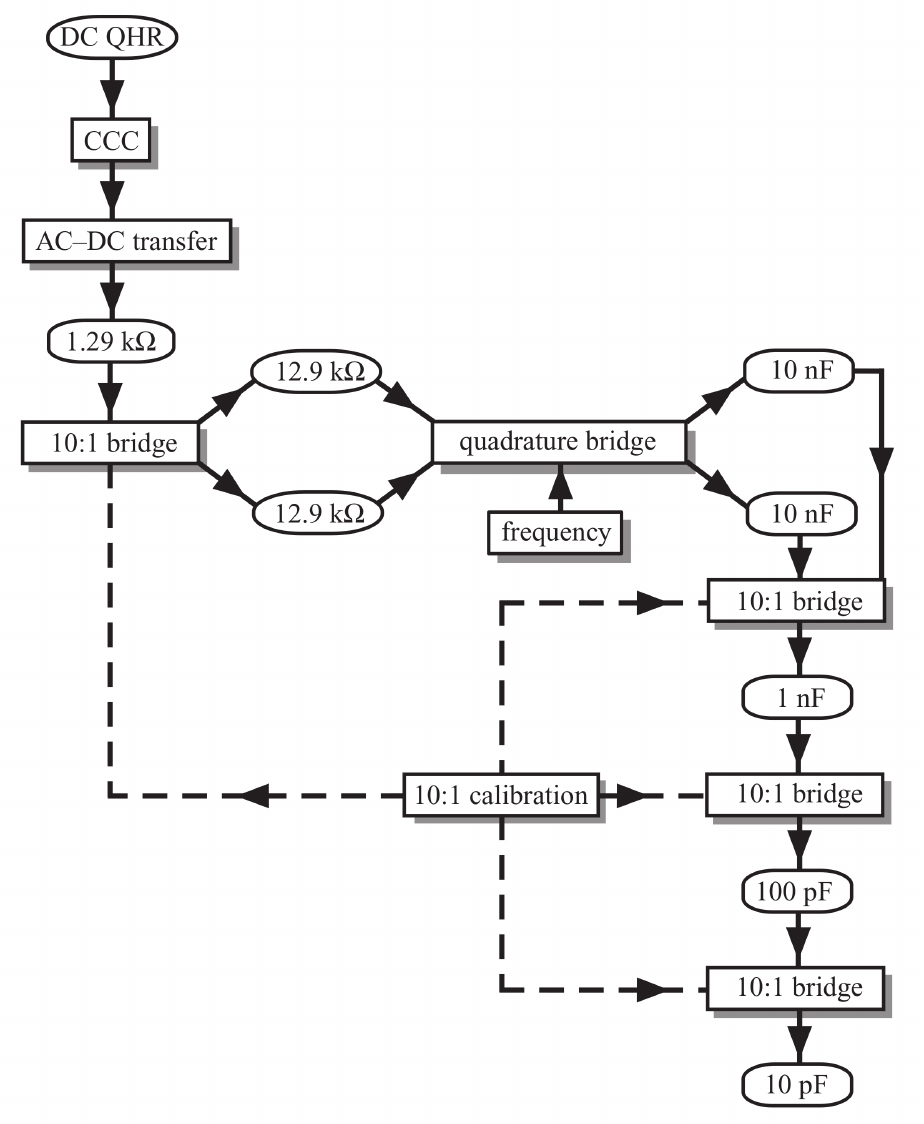}
	\caption[Capacitance traceability chain]{Traceability chain for the realisation of capacitance (at $\SI{10}{\pico\farad}-\SI{100}{\pico\farad}$ level) from the dc quantum Hall effect. After Ref.~[Sec.~6.37]\cite{Awan:2010}}
	\label{fig:tradchain}
\end{figure}
%
%
In primary electrical impedance metrology a traceability chain starts with the realisation of the ohm from the quantum Hall effect.

The chain then logically\footnote{The ordering in time of the comparisons can be different from the logical ordering implied in the traceability chain.} proceeds by comparisons performed with dedicated measurement setups, until the calibration of a set of impedances (maintained standard) of very high stability is achieved. 

The typical target of a capacitance traceability chain is the calibration of capacitors of \SI{10}{\pico\farad} or \SI{100}{\pico\farad} nominal values, since these are the most stable available and hence the capacitance unit can be maintained between successive calibration at the highest accuracy level.
\subsection{Traditional traceability chain}
An example of a traditional traceability chain from the (dc) quantum Hall effect to \SI{10}{\pico\farad} capacitance is shown in Figure~\ref{fig:tradchain}. Others can be found in  \cite{Nakamura:2001,Callegaro:2010, Kim:2019}. 

In Figure~\ref{fig:tradchain} the DC quantum Hall resistance is employed to calibrate in dc, using a cryogenic current comparator (CCC), an ac-dc resistor of appropriate value (in this case \SI{1.29}{\kilo\ohm} of calculable frequency performance~\cite{Haddad:1969, Gibbings:1963, Bohacek:2001}. A resistance transformer ratio bridge is employed to scale up in the ac regime the resistance, calibrating two resistance standards $R_1$ and $R_2$ (in this case of \SI{12.9}{\kilo\ohm}. A quadrature bridge can calibrate the product of two capacitors $C_1$ and $C_2$ under the condition $\omega^2 R_1 R_2 C_1 C_2 = 1$; in this example, two \SI{10}{\nano\farad} capacitors at the frequency of about \SI{1233}{\hertz}. Further scaling with a capacitance transformer ratio bridge allows to realise a capacitance scale down to \SI{10}{\pico\farad}.
\subsubsection{GIQS traceability chains}
\begin{figure}[p]
\begin{minipage}[b]{0.46\textwidth}
\centering
\includegraphics[width=0.7\textwidth,clip=]{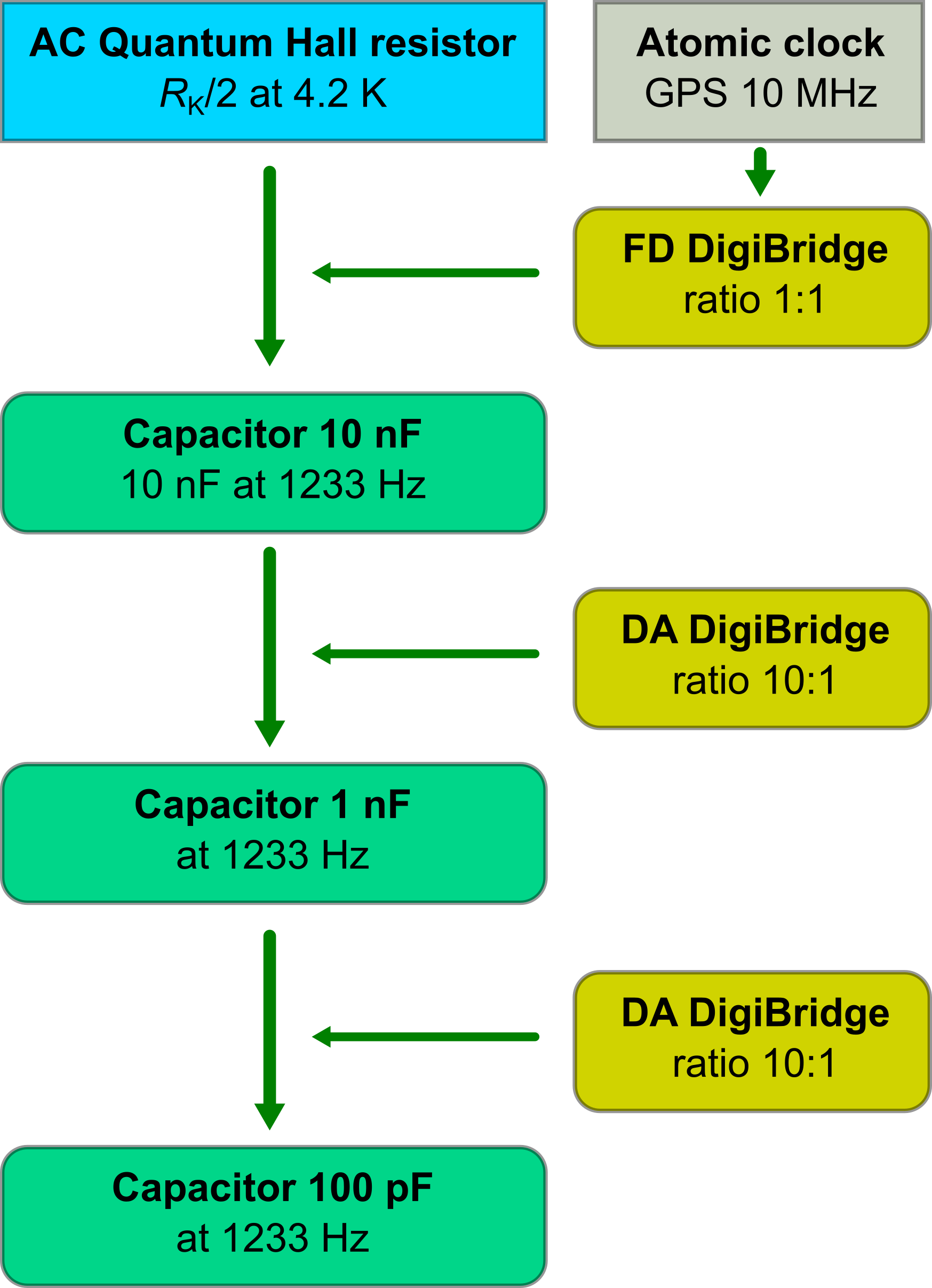}
\end{minipage}
\hfill
\begin{minipage}[b]{0.46\textwidth}
\centering
\def\svgwidth{0.65\textwidth}
{\footnotesize 
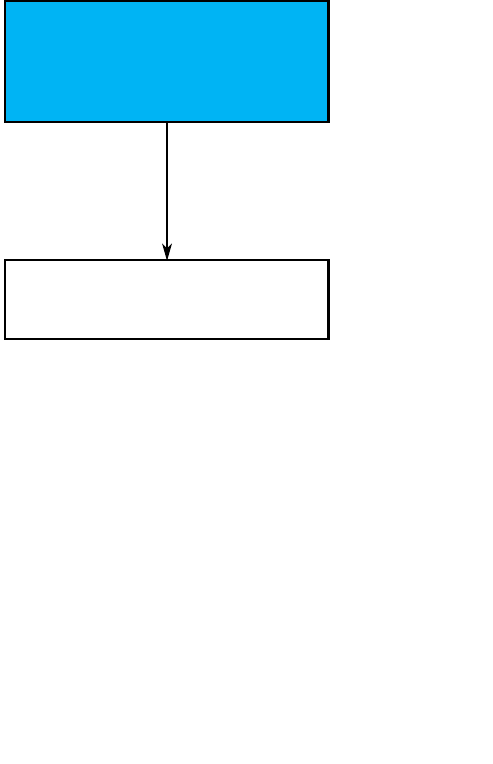}
\end{minipage}

\vspace{2ex}
\begin{minipage}[t]{0.46\textwidth}
\centering 
(a)
\end{minipage}
\hfill
\begin{minipage}[t]{0.46\textwidth}
\centering 
(b)
\end{minipage}

\vspace{2ex}
\begin{minipage}[b]{0.92\textwidth}
\includegraphics[width=\textwidth,viewport=51 17 707 409,clip=]{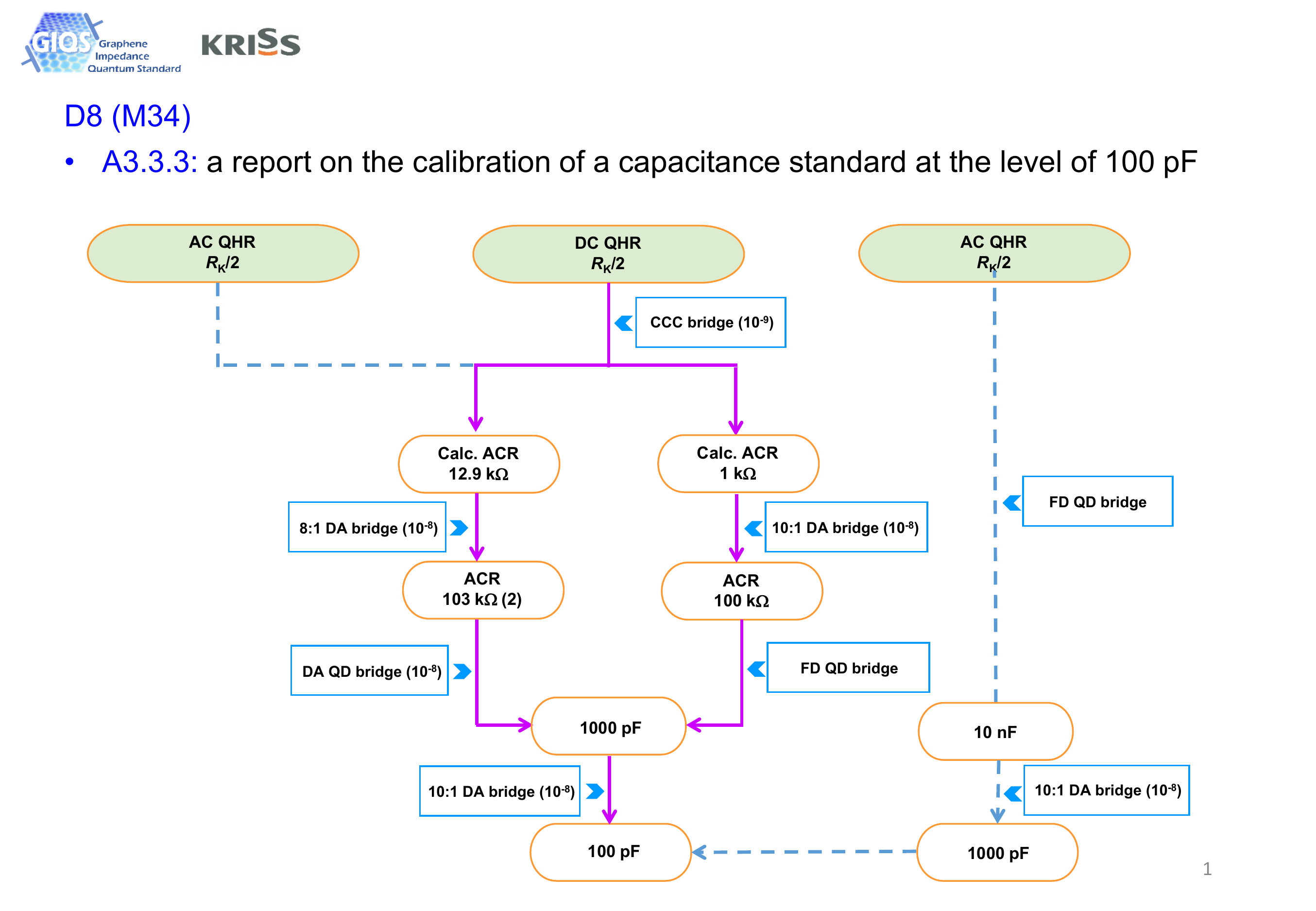}
\end{minipage}

\vspace{2ex}
\begin{minipage}[t]{0.92\textwidth}
\centering 
(c)
\end{minipage}
\caption[CMI, INRIM, KRISS capacitance traceability chains]{Traceability chains under development at (a) CMI, (b) INRIM and (c) KRISS within the framework of the project 18SIB07 GIQS, from a graphene AC QHR standard to a \SI{100}{\pico\farad} standard capacitor.}
\label{fig:traceability_chains}
\end{figure}
The GIQS project results allow to consideraby simplify the traditional traceability chain of Figure~\ref{fig:tradchain} via three routes:
\begin{description}
\item[ACQHE] The direct exploitation of the quantum Hall resistance in the AC regime (ACQHR) allows to avoid completely the need of a calculable resistor to perform the AC-DC resistance transfer~\cite{Inglis:2003,Schurr:2009, Schurr:2012}. The calculable resistors and the associated DC measurements can be avoided;
\item[$R-C$ transfer] The most complex step of the traditional traceability chain is the $R-C$ comparison, which asks for four standards. As described in Section~\ref{sec:elecfdbridges} the $R-C$ transfer with digital bridges can be performed between two standards, a resistor and a capacitor
\item[$C$ scaling] Josephson bridges (Section~\ref{sec:josephsonbridges}) allow to perform ratio measurement for capacitance scaling without the need of a ratio standard calibration, as occurs for the inductive voltage divider of traditional transformer bridges. Electronic digitally-assisted bridges can also considerably simplify the task of ratio measurements, although they still require a ratio calibration.
\end{description}
As examples, we briefly summarize three traceability chains developed within GIQS:
\begin{description}
\item[CMI] The CMI traceability chain, figure~\ref{fig:traceability_chains}(a), starts with the calibration at \SI{1233}{\hertz} of a \SI{10}{\nano\farad} capacitance standard against a graphene ACQHR standard by means of an electronic fully-digital quadrature bridge, working at $1:1$ impedance magnitude ratio. Then, the \SI{10}{\nano\farad} value is scaled down to a \SI{1}{\nano\farad} capacitance standard by means of a $10:1$ digitally-assisted bridge operating again at \SI{1233}{\hertz}. Finally, the \SI{1}{\nano\farad} value is scaled down to a \SI{100}{\pico\farad} capacitance standard with the same digitally-assisted bridge.
\item[INRIM] The INRIM traceability chain, figure~\ref{fig:traceability_chains}(b), starts with the calibration at \SI{1541}{\hertz} of an \SI{8}{\nano\farad} solid-dielectric capacitance standard against a graphene ACQHR standard by means of an electronic quadrature fully-digital bridge~\cite{Marzano:2020}, working at $1:1$ impedance magnitude ratio. Then, the \SI{8}{\nano\farad} value is scaled down to a \SI{1}{\nano\farad} gas capacitance standard by means of an $8:1$ digitally-assisted bridge~\cite{Callegaro:2010} operating again at \SI{1541}{\hertz}. Finally, the \SI{1}{\nano\farad} value is scaled down to a \SI{100}{\pico\farad} quartz capacitance standard by means of a capacitance build-up method~\cite{Tran:2020}.
\item[KRISS] is evaluating different possible traceability chains as shown in figure~\ref{fig:traceability_chains}(c). The leftmost one is the current traceability chain starting with a DCQHR standard and which was described in~\cite{Kim:2019}. The rightmost chain starts with an ACQHR standard and is the same CMI chain described above. The center chain starts with a DCQHR standard which is used to calibrate a \SI{1}{\kilo\ohm} calculable AC/DC Haddad resistance standard by means of a cryogenic current comparator. The \SI{1}{\kilo\ohm} standard is then used to calibrate a \SI{100}{\kilo\ohm} AC resistance standard by means of a $10:1$ digitally-assisted bridge, in two successive steps. The \SI{100}{\kilo\ohm} resistor value is transferred to a \SI{1}{\nano\farad} capacitance standard by means of an electronic quadrature fully-digital bridge. Finally, the \SI{1}{\nano\farad} value is scaled down to a \SI{100}{\pico\farad} capacitance standard with the $10 : 1$ digitally-assisted bridge.

\end{description}

\subsection{Capacitance artifact standards}
National metrology institutes maintain a local capacitance scale, typically composed of one or more standard for each decadal value. Each standard is thoroughly characterized for the influence of the environmental parameters (typically temperature, sometimes humidity and temperature also) and calibrated periodically; its value is monitored over the long term to determine the drift over time, so a prediction of the capacitance value and the corresponding in-use uncertainty at any time after the last calibration can be obtained. 
The range covered by commercial capacitance standards starts from \SI{1}{\pico\farad} and goes up to \SI{100}{\micro\farad}. Higher values, up to \SI{1}{\farad}, are synthesized by passive transformer standards \cite{Hall:1976,Kuperman:2010} or with the aid of electronic amplifiers.

\subsubsection{Fused-silica capacitors}
The most stable capacitance standards, used for long-term maintenance of the farad unit in national metrology institutes,  are given by monolithic fused-silica or fused-quartz standards, available in the \SI{1}{\pico\farad} to \SI{100}{\pico\farad} range. The present construction is due to Cutkosky \cite{Cutkosky:1965}: active and shield silver film electrodes are directly fired over a dielectric ``hockey puck'' and the element contacted by springs in a supporting cell.

The frequency dependence of fused-silica capacitors is small (of the order of \num{1E-6} in the audio frequency range~\cite{Wang:2003}) and the temperature coefficient is around \SI{12E-6}{\per\kelvin} \cite{Daniel:1995}. The time drift can be below \SI{E-7} per year. A commercial model with a thermostated enclosure is available (Andeen-Hagerling AH1100) is available and has been employed in international intercomparisons~\cite{Gournay:2018}.

\subsubsection{Gas-dielectric capacitors}
The range from \SI{1}{\pico\farad} to \SI{10}{\nano\farad} can be also covered by sealed gas-dielectric capacitors~\cite{Hersh:1963, Dunn:1964}. Gas-dielectric capacitors can have a low (less than \SI{2E-6}{\per\kelvin}~\cite{Hersh:1963}) temperature coefficient, and a very low (in the \num{E-6} range) loss. The very small frequency dependence in the audio frequency range,can be evaluated from radio-frequency measurements~\cite{Callegaro:2003S, Callegaro:2008HF}.

\subsubsection{Solid-dielectric capacitors}
Standards from \SI{1}{\nano\farad} to \SI{100}{\micro\farad} are available as dielectric (mica, polymer) capacitors. The temperature dependence (several \SI{E-5}{\per\kelvin}), the frequency dependence and loss (in the \num{E-5}-\num{E-4} range) is much larger than silica or gas-dielectric standards, and the time stability is typically worse. Standards made from electronic components in a thermostated enclosure have been realized~\cite{NPLstdcap, Callegaro:2005, Xiaobing:2016}
\subsection{Calibration of commercial meters}
\subsubsection{Capacitance meters}
At the present time the highest-accuracy capacitance meters available on the market are the Andeen-Hagerling AH2500A, AH2550A (fixed frequency, \SI{1}{\kilo\hertz} and AH2700A (variable frequency, \SI{20}{\hertz} to \SI{20}{\kilo\hertz}). The measurement range is up to \SI{1}{\micro\farad}. The bridges rely on internal fused-silica capacitance standards; an artifact calibration mode is available, performed by measurement of a capacitor in the \SI{10}{\pico\farad} to \SI{1600}{\pico\farad} range.
\subsubsection{Impedance meters}
In commercially available impedance meters (also called $LCR$ or $RCL$ meters/bridges, or impedance analyzers) the ratio standard necessary for the measurement action is provided by an electronic circuit, either analog or digital. The impedance being measured is compared with a internal reference impedance (chosen within a set). The base accuracy is typically limited to parts in \num{E4}. 

The calibration of impedance meters is performed with a set of impedance standards of different kind (resistors, capacitors and inductors) and different nominal values. Each standard has to be manually connected to the meter. The difference between the reading value and the reference value (the latter coming from a calibration certificate) is the meter reading error, for that particular nominal impedance value and measurement frequency. Often, electronic impedance bridges allow to perform the so-called \emph{artifact calibration}, which is a \emph{de facto} adjustment procedure: for each impedance standard measured, the corresponding reference numerical value is entered (typically with a keyboard) in the bridge memory. A firmware calculates a set of adjustment numerical coefficients and store them in the bridge permanent memory. The coefficients are then employed during normal measurements, to convert raw data into readings.
%
%
\bibliographystyle{IEEEtran}
\bibliography{A415_GPG}
\end{document}